\documentclass[12pt]{article}
\usepackage{jheppub}
\pdfoutput = 1

\usepackage{amsmath}
\usepackage{amssymb}
\usepackage{bbm}
\usepackage{bbold}
\usepackage{braket}
\usepackage{caption}
\usepackage{color}
    \definecolor{darkgreen}{rgb}{0,0.5,0}
    \definecolor{darkred}{rgb}{0.5,0,0}
    \definecolor{darkblue}{rgb}{0,0,0.6}
    \definecolor{purple}{rgb}{0.4,.2,0.7}

\usepackage{float}
\usepackage{graphicx}
\usepackage{hyperref}

\usepackage{mathabx}
\usepackage{mathtools}
\usepackage{pdfsync}
\usepackage{slashed}
\usepackage[normalem]{ulem}
\usepackage{upgreek}
\usepackage{url}

\usepackage[ragged]{footmisc}
\def\be{\begin{equation}}
\def\ee{\end{equation}}

\renewcommand{\tilde}{\widetilde}

\numberwithin{equation}{section}

\begin{document}
%\subheader{empty}
\title{Almost all extremal black holes in AdS are singular}

\author[a]{Gary~T.~Horowitz,}
\author[b]{Maciej Kolanowski,}
\author[c]{Jorge~E.~Santos}
%affilations
\affiliation[a]{Department of Physics, University of California at Santa Barbara, Santa Barbara, CA 93106, U.S.A.}
\affiliation[b]{Institute of Theoretical Physics, Faculty of Physics, University of Warsaw, Pasteura 5, 02-093 Warsaw, Poland}
\affiliation[c]{Department of Applied Mathematics and Theoretical Physics, University of Cambridge, Wilberforce Road, Cambridge, CB3 0WA, UK}

% e-mail addresses
\emailAdd{horowitz@ucsb.edu}
\emailAdd{maciej.kolanowski@fuw.edu.pl}
\emailAdd{jss55@cam.ac.uk}

\abstract{We investigate the geometry near the horizon of a generic, four-dimensional extremal black hole.   When the cosmological constant is negative, we show that (in almost all cases )  tidal forces diverge as one crosses the horizon, and this singularity is stronger for larger black holes. In particular, this applies to generic nonspherical black holes, such as those satisfying inhomogeneous boundary conditions. Nevertheless, all scalar curvature invariants remain finite. Moreover, we show that nonextremal black holes have tidal forces that diverge in the extremal limit. Holographically, this singularity is reflected in anomalous scaling of the specific heat with temperature.
Similar (albeit weaker) effects are present when the cosmological constant is positive, but not when it vanishes.}

\maketitle
%%%%%%%%%%%%%%%%%%
\section{Introduction} 

In four-dimensional general relativity, asymptotically flat, stationary black holes have extremal limits with smooth horizons. This follows from the black hole uniqueness theorems and the known properties of the Reissner-Nordstr\"om and Kerr solutions. Over the years, various examples have been found showing that this is not always the case. A mild lack of smoothness (where the metric is $C^2$ but not $C^3$)  was first noticed in the static multi-black hole solutions to $D=5$ Einstein-Maxwell theory \cite{Welch:1995dh}. This became more serious with the discovery that in $D>5$, static multi-black hole solutions have curvature singularities on the horizon \cite{Candlish:2007fh}. These were null singularities in which tidal forces on infalling observers diverge, but all curvature scalars remain finite.

Similar null singularities were also seen in the extremal limit of some black holes in anti-de Sitter (AdS) space. This includes solutions with less symmetry \cite{Dias:2011at,Maeda:2011pk, Hickling:2015ooa}, nonsupersymmetric attractor flows \cite{Iizuka:2022igv} and even in some supersymmetric black holes \cite{Markeviciute:2018yal}. A natural question to ask is how common are these singular extremal solutions? Are they exceptional special cases, or indicating a more general phenomenon? 

We will show that in AdS they are very common. In fact, almost all extremal black holes are singular. This is true even in four dimensions (and becomes worse in higher dimensions). We will focus on four dimensional solutions of Einstein-Maxwell theory with $\Lambda < 0$. The higher dimensional case will be discussed elsewhere \cite{Horowitz:2022leb}. There are many more stationary black holes in AdS than in asymptotically flat spacetime since one has the freedom to choose boundary conditions for the metric and vector potential at infinity. In particular, static nonspherical charged black holes exist,  but we will show they are generically singular.   Our results apply whenever rotational symmetry is broken, so for example, if one puts a cage around a static AdS black hole it becomes singular. 

If the horizon was smooth, it is known that in the extremal limit, the only possible static near horizon geometry is $AdS_2 \times S^2$ \cite{Kunduri:2008tk} so the horizon itself remains spherical. We will see that in four dimensions, even when the horizon becomes singular, a well defined near horizon geometry exists and remains  $AdS_2 \times S^2$. Intuitively, this is because
the radial distance (along a static hypersurface) from the horizon to any point outside is infinite and thus  any nonspherical perturbation\footnote{One should note that in this work  a ``perturbation" does not mean any dynamical change, but rather a change in boundary conditions for the elliptic problem of finding static black holes. In particular, our work is different from the Aretakis instability of extremal black holes \cite{Aretakis:2011ha} which results from time dependent perturbations (although both effects originate from the symmetries of the near horizon $AdS_2$ factor).}  should decay before reaching the horizon. But the key point is how quickly do they decay. The symmetry of $AdS_2$ ensures that all perturbations should have power law behavior near an extremal horizon. If the exponent is not an integer, the solution is not $C^\infty$, and if the exponent is too small, the curvature will diverge. We will show that for AdS black holes with topology $S^2$, an $\ell =2$ perturbation always falls off slowly enough to produce a singularity on the horizon. So generic extremal black holes with $S^2$ topology are singular. This singularity is null, and all curvature scalars remain finite. However, the tidal forces on infalling particles diverge.

As one increases the charge, this singularity becomes stronger and higher $\ell$ modes also become singular. Similar results hold 
for static black holes of different topologies (with the exception of small toroidal ones) and for Kerr-AdS. In fact, for large hyperbolic black holes, the singularity is so strong that some perturbations  diverge at the horizon. Thus, we see that almost all extremal black holes in AdS are singular. Smoothness of the known exact solutions is an artifact of the symmetry rather than a basic physical feature. Solutions with extremal AdS black holes in nonspherical backgrounds have been constructed before \cite{Horowitz:2014gva}. Although it was not noticed at the time, the current analysis shows that these ``hovering" black holes also have diverging tidal forces on their horizon.

A natural question is whether our assumption that the cosmological constant is negative is needed at all. Even when $\Lambda \ge 0$, if the horizon was smooth, the only static near horizon extremal black hole geometry would be $AdS_2\times S^2$, so all perturbations must fall off like a power law. However when $\Lambda = 0$, one finds that all  exponents are positive integers and so the metric is indeed smooth. In this case, since there is no other scale and the exponents are dimensionless, they cannot depend on the charge. They turn out to be integers in four dimensions, but not in higher dimensions.    When $\Lambda$ is positive, the exponents are no longer integers and small black holes are singular.

The program of systematic investigations of the spacetime near  extremal horizons was first proposed in  \cite{Li:2015wsa} and was continued in \cite{Fontanella:2016lzo, Li:2018knr, Kolanowski:2019wua, Kolanowski:2021tje}. Unfortunately, the starting point of that analysis was the Taylor expansion in the distance from the horizon. This clearly assumes smoothness and is generically not allowed. Thus, one should rather see these results (at least with $\Lambda \neq 0$) as a search for very special, fine-tuned solutions. This clarifies the conclusion of \cite{Kolanowski:2021tje} where it was shown that the transversal deformation of the extremal Reissner--Nordstr\"om-(A)dS horizon are spherically symmetric unless the charge takes a special value (depending on the cosmological constant).

If one considers a nonextremal black hole with temperature $T$, these singularities are always removed. Thus, it is tempting to simply ignore them as an artifact of $T \to 0$ limit. However, as we will show, even in this case tidal forces at the horizon grow as an inverse power of the temperature and diverge in the limit. Thus even a tiny, symmetry breaking perturbation at infinity becomes arbitrarily large near the horizon as we lower $T$. This large curvature may lead to quantum corrections near the horizon, but   we do not currently understand the form of these corrections. In a holographic theory, we will show that there is a clear signal of the singularity for large black holes: the specific heat (and other quantities) has anomalous scaling with $T$ near $T=0$.\footnote{We thank Sean Hartnoll for suggesting this might occur.} 

The reason these singularities exist and some of their properties can already be seen by looking at a massless scalar field in an extremal black hole background. 
So we start by discussing this simple example in the next section. In Sec.  3, we begin our main analysis of Einstein-Maxwell solutions, by studying linearized gravitational and electromagnetic perturbations of the near horizon geometry of extremal black holes. Sec. 4 contains a discussion of the full nonlinear story, and shows that the singularities indicated by the linearized analysis indeed arise in the full solutions as $T\to 0$. To see the anomalous scaling of the specific heat, one needs to go to very low $T$, which is difficult to reach in the Einstein-Maxwell theory. So in Sec. 5 we introduce a simpler theory  in which this effect can be clearly demonstrated. We conclude in Sec. 6 with a brief discussion.

%%%%%%%%%%%%%%%%%%%%%%%%
\section{\label{sec:simpe}Simple example}

Before we get into the technical details, let us consider a very simple toy model which will illustrate the main ideas. We will consider a massless scalar field on an extremal Reissner--Nordstr\"om-AdS (RN ASdS) black hole. 
Recall that the RN AdS metric is
\begin{equation}
       {\rm d}s^2 = - f(r)\,{\rm d}t^2 + \frac{{\rm d}r^2}{f(r)} + r^2 {\rm d}\Omega^2
\end{equation}
where  $ {\rm d}\Omega^2$ is the line element on a unit radius round two-sphere,
\be
f(r) = \frac{r^2}{L^2} + 1-\frac{2M}{r} +\frac{Q^2}{r^2}
\ee
and $L$ is the AdS radius. 
In the extremal limit, the horizon is at
\be\label{radiusQ}
 r_+ = \sqrt{\frac{2Q^2}{1 + \sqrt{1 +12 {Q^2}/{L^2}}}}
 \ee
and 
\be
f''(r_+) = \frac{6}{L^2} + \frac{2 Q^2}{r_+^4}
\ee

We now perturb this spacetime by adding a static, massless scalar field $\phi$. Since the background is spherically symmetric, we may expand $\phi$ into the spherical harmonics:
\begin{equation}
    \phi = \sum_{\ell, m} \phi_{\ell m} Y_{\ell m}.
\end{equation}
Then, the Klein--Gordon equation reads
\be
(f\phi_{\ell m}')' + \frac{2f\phi_{\ell m}'}{r} - \frac{\ell(\ell+1)}{r^2} \phi_{\ell m} =0
\ee
This is a simple ODE with $r=r_+$ being a regular singular point. Thus, near $r=r_+$ we can approximate it by the Euler equation:
\begin{equation}
     \frac{1}{2}(r-r_+)^2  f''(r_+) \phi_{\ell m}'' +  (r-r_+)  f''(r_+) \phi_{\ell m}' - \frac{\ell (\ell+1)}{r_+^2} \phi_{\ell m} = 0,
\end{equation}
and so near the horizon we have $\phi_{\ell m} \sim (r-r_+)^{\gamma_\pm}$, where
 \begin{equation}
 \gamma_\pm=\frac{1}{2}\left[
 \pm \sqrt{1+\frac{4\ell(\ell+1)}{1+6\,r_+^2/L^2}}-1\right]\,, \label{eq:n_KG}
% \label{eq:n}
 \end{equation}
 %where we defined the dimensionless radius $y_+\equiv r_+/L$.

We, of course, choose $\gamma_+$ since the other choice would lead us to a highly singular solution. Nevertheless, notice that when $\ell=1$, we have $0<\gamma_+<1$ for all $r_+>0$. Thus, the field is only $C^0$ at the horizon. Moreover, certain components of the associated energy--momentum tensor:
\begin{equation}
    T_{rr} \sim (\phi_{,r})^2 \sim (r-r_+)^{2(\gamma_+ - 1)}
\end{equation}
are divergent, so the backreaction on the metric will produce a singularity. Nevertheless, all scalar quantities built from $T_{\mu \nu}$, such as $T$ or $T_{\mu \nu} T^{\mu \nu}$ are finite. Thus, one could be tempted to blame our choice of coordinates for the apparent singularity. However, that would be not justified since $r$ is a good coordinate at the horizon. Moreover, if one replaces $t$ with an ingoing Eddington coordinate $v$, the calculation is the same, and now  $\partial_r$ is a vector field tangent to the affinely parametrized null geodesics and so it has a clear geometrical meaning. Also in this case, $T_{rr}$ enters the Raychaudhuri equation and so its divergence signifies that the family of null rays emanating from the horizon is singular. This is going to be a general lesson for all the examples we consider later in this paper: generic nonspherical perturbations produce a physical curvature  singularity along the null horizon, but all curvature scalars are finite. Thus, one needs to be extra careful with the choice of coordinates to properly capture these divergences. 

A few remarks are in order regarding \eqref{eq:n_KG}: 
\begin{itemize}
    \item Although we assumed that $\phi$ is massless, similar conclusions would hold also for massive but light fields. Thus, it is not just a result of an unfortunate fine-tuning of the  model.
    \item If $r_+/L$ is large enough, $\gamma_+ < 1$ also for higher $\ell$'s. 
    \item The larger  $r_+$ is, the smaller $\gamma_+$ and thus the solutions are more and more divergent. As we will see, this and the previous remark hold also in the nonlinear Einstein--Maxwell theory. This means that (counter-intuitively) large black holes, whose curvature scalars at the horizon are much less than small black holes, nevertheless  have stronger singularities if we perturb them a little bit.
    \item Eq.~\eqref{eq:n_KG} does not depend on any asymptotic conditions. It was derived locally, just near the horizon. The only role of the asymptotic region is to provide a source for non-symmetric modes.
    \item The case $\Lambda = 0$ can be read off from Eq.~\eqref{eq:n_KG} by taking $L\to \infty$. The result is $\gamma_+ = \ell$, so $\phi$ remains smooth.
    \item The case  $\Lambda > 0$ can be read off from Eq.~\eqref{eq:n_KG} by analytically continuing $L^2 \to -L^2$. One sees that $\phi$ is at least $C^1$ but it is still not smooth. As we will see, for {\it certain} black holes in dS, the singularity at the horizon may persists, although it will be milder.
\end{itemize}

Eq. \eqref{eq:n_KG} can be understood as a special case of a familiar result in gravitational holography. The near horizon geometry of the extremal RN AdS solution is $AdS_2\times S^2$ with $AdS_2$ radius $L_2 = [2/f''(r_+)]^{1/2}$. The $\ell^{th}$ harmonic acts just like a field of mass $m^2 = \ell(\ell+1)/r_+^2 $ in this $AdS_2$ spacetime. In terms of $L_2$ and $m^2$, eq. \eqref{eq:n_KG} becomes
\be\label{eq:scalingdim}
\gamma_\pm = \frac{ -1 \pm \sqrt{1 + 4m^2 L_2^2}}{2}
\ee
This is a special case of a more general formula  that gives the power law behavior of fields with mass $m$ in $AdS_D$, which is the scaling dimension of the dual operator.\footnote{Eq. \ref{eq:scalingdim} differs from the usual scaling dimension by an overall sign, since we have defined it to be the power of $r-r_+$, rather than the more commonly used power of an inverse radius.}

\section{Einstein--Maxwell: linear theory}
\subsection{General equations}
We consider the following equations of motion
\begin{subequations}
\label{EoMs}
\begin{equation}
    R_{\mu \nu} = 2 F_{\mu \sigma} F_{\nu}^{\ \sigma} - \frac{1}{2} g_{\mu \nu} F_{\alpha \beta} F^{\alpha \beta} - \frac{3}{L^2} g_{\mu \nu},
\end{equation}
\begin{equation}
    \mathrm{d} F = 0,
\end{equation}
\begin{equation}
    \mathrm{d} \star F = 0,
\end{equation}
\end{subequations}
where $F = {\rm d}A$ is the Maxwell two-form, $A$ is its potential and $L$ is the radius of $AdS_4$.

We are interested in the solutions to \eqref{EoMs} which describe a stationary extremal black hole. Near the horizon we may introduce Gaussian null  coordinates $(v,\rho,x^a)$ in which the metric and Maxwell field read:
\begin{subequations}
\label{gaussian}
\begin{equation}
    g = 2{\rm d}v \left( {\rm d}\rho + \rho\,h_{a}\,{\rm d}x^a - \frac{1}{2}\,\rho^2\,C\,{\rm d}v \right) + q_{ab}\,{\rm d}x^a\,{\rm d}x^b
\end{equation}
\begin{equation}
    F = E\,{\rm d}v \wedge {\rm d}\rho + \rho\,W_{a}\,{\rm d}v \wedge {\rm d}x^a + Z_a\,{\rm d}\rho \wedge {\rm d}x^a + \frac{1}{2}\,B_{ab}\,{\rm d}x^a \wedge {\rm d}x^b,
\end{equation}
\end{subequations}
where nothing depends on $v$ (so $\partial_v$ is the Killing vector generating the horizon). It is often useful to work with the near horizon geometry of the spacetimes of the form \eqref{gaussian}. To this end we consider a one-parameter $(\epsilon > 0)$ family of diffeomorphisms
\begin{equation}
    \phi_\epsilon (v,\rho,x^a) = (\epsilon^{-1} v, \epsilon \rho, x^a).
\end{equation}
The limits of pull-backs
\begin{equation}
    \lim_{\epsilon \to 0} (\phi_\epsilon^\star g, \phi_\epsilon^\star F) = (\mathring{g}, \mathring{F})
\end{equation}
exist and provides us with a new smooth solution to the Einstein--Maxwell equations. Then, \eqref{gaussian} simplifies significantly
\begin{subequations}
\label{nhg_gaussian}
\begin{equation}
    \mathring{g} = 2\,{\rm d}v \left( {\rm d}\rho + \rho\,h_{a}\,{\rm d}x^a - \frac{1}{2}\,\rho^2\,C {\rm d}v \right) + q_{ab}\,{\rm d}x^a\,{\rm d}x^b
\end{equation}
\begin{equation}
    \mathring{F} = E\,{\rm d}v \wedge {\rm d}\rho + \rho\,W_{a}\,{\rm d}v \wedge {\rm d}x^a + \frac{1}{2} \,B_{ab}\,{\rm d}x^a\,\wedge {\rm d}x^b,
\end{equation}
\end{subequations}
where now all the $\rho$-dependence is explicit. Notice that $\mathring{g}$ posses a new Killing vector: $\rho\partial_\rho -v\partial_v $.  Also, \eqref{EoMs} simplifies significantly for $(\mathring{g}, \mathring{F})$. This allowed for the classification (under the assumption of smoothness and either  staticity or axial symmetry) of geometries of the extremal horizons in four dimensions. The only possible geometries are either those of Reissner--Nordstr\"om-(AdS) or Kerr--Newman--(AdS) (in the static or the axially symmetric case, respectively).
Below we consider how stationary solutions to \eqref{EoMs} behave near those horizons.

Since we are interested only in the near horizon behavior, we may write our (generic yet stationary) fields as
\begin{subequations}
\begin{equation}
    g = \mathring{g} + \delta g,
\end{equation}
\begin{equation}
    F = \mathring{F} + \delta F
\end{equation}
\end{subequations}
where $(\delta g, \delta F)$ are supposed to vanish on the horizon (and by continuity, are small nearby). Thus, it seems reasonable to expect that $(\delta g, \delta F)$ satisfies {\it linearized} Einstein--Maxwell equations on the background of $(\mathring{g}, \mathring{F})$. Due to the symmetries, we may decompose our perturbations into eigenspaces of $\rho \partial_\rho - v \partial_v$. They are thus of the form
\begin{subequations}
\label{perturbation}
\begin{equation}
    \delta g = \rho^\gamma \left(
    \delta F\,\rho^2\,{\rm d}v^2 + 2\,\rho\,\delta h_a\,{\rm d}v\,{\rm d}x^a + \delta q_{ab}\,{\rm d}x^a\,{\rm d}x^b
    \right) 
    \end{equation}
    \begin{equation}
    \delta\mathcal{F} = \rho^\gamma \left(
    \delta E\,{\rm d}v\wedge {\rm d}\rho + \rho\,\delta W_a\,{\rm d}v\wedge {\rm d}x^a + \rho^{-1} \delta Z_a\,{\rm d}\rho \wedge {\rm d}x^a + \frac{1}{2}\,\delta B_{ab}\,{\rm d}x^a \wedge {\rm d}x^b
    \right).
    \end{equation}
\end{subequations}
The scaling symmetry implies
\begin{subequations}
\begin{equation}\label{singweyl}
    \delta C_{\rho a \rho b} \sim \gamma (\gamma-1) \rho^{\gamma-2}
\end{equation}
\begin{equation}
    \delta R_{\rho \rho} \sim \gamma (\gamma-1) \rho^{\gamma-2},
\end{equation}
\end{subequations}
{where $C_{\alpha\beta\mu\nu}$ is the Weyl tensor.} 
Thus, we see that if $1 \neq \gamma<2$, then our linearized solutions are singular. We will show that there are indeed solutions  with $0<\gamma<1$ which strongly suggests that generically the spacetime is singular at the horizon.\footnote{If $0<\gamma < 1$, the metric is continuous, but not differentiable at the horizon. Nevertheless, it still makes sense to ask if the norm of the timelike Killing field has a double zero at the horizon (so extremal black holes are well-defined) since this function remains $C^2$.
 Weak solutions of the Einstein--Maxwell equations can be defined if the curvature is integrable. This requires the Christoffel symbols to be square integrable. If $\gamma>\frac{1}{2}$ this is the case, so one can extend the fields inside the horizon as a weak solution, but the extension is not unique.  As we will see, for sufficiently large black holes,  that condition is not satisfied and no extension is possible.}

One might hope that our results are just an artifact of the linearized approximation and the full nonlinear solution would behave differently. However, this is not the case. Even though the curvature diverges, the metric perturbation is small near the horizon, so higher order corrections to the metric will be even smaller. One can show that the same scaling results hold in the full theory, as long as fields fall off like a power law near the horizon. In Sec. \ref{sec_finite_temp}, we will verify (numerically) that this is indeed the case for asymptotically AdS black holes. For the rest of this section, we will determine $\gamma$ using a linearized analysis.

Notice that since the diverging components always involve a $\rho$ index and the inverse of the metric in \eqref{nhg_gaussian}  has $\mathring{g}^{\rho \rho} =-C \rho^2$, all curvature invariants will remain finite at the horizon.

\subsection{Reissner--Nordstr\"om--AdS\label{sec:per}}

As we mentioned above,
the only static near horizon geometry is given by the limit of the extremal Reissner--Nordstr\"om-AdS solution. Since in  four dimensions there is a duality between electric and magnetic fields, we may assume that our black hole has only an electric charge. The fields simplify significantly and they read:
\begin{subequations}
\begin{equation}
    \mathring{g} = 2\,{\rm d}v \left( {\rm d}\rho - \frac{1}{2}\,\rho^2\,C\,{\rm d}v \right) + q_{ab}\,{\rm d}x^a\,{\rm d}x^b
\end{equation}
\begin{equation}
    \mathring{F} = E\,{\rm d}v \wedge {\rm d}\rho,
\end{equation}
\end{subequations}
where now $C,E$ are constants and $q$ is a two-dimensional metric of constant curvature. The first term is just $AdS_2$ with a length scale set by $C$. The field equations require:
\begin{subequations}
\begin{equation}
R = -\frac{6}{L^2} + 2E^2 \label{background_parameters}
\end{equation}
\begin{equation}
C = \frac{3}{L^2} + E^2,
\end{equation}
\end{subequations}
where $R$ is the Ricci scalar of $q$. In short, this says that the near horizon solution has a product structure $AdS_2 \times H$, where $H$ has a constant curvature (of any sign). 

Although it is possible to derive and solve equations for the ansatz \eqref{perturbation}, it is not the most convenient way to find exponents $\gamma$. Indeed, we perturb a highly symmetrical background so one should take an advantage of that. We may thus decompose $\delta g$ and $\delta F$ into the eigentensors of the Laplacian on $H$ and then use the Kodama--Ishibashi formalism. Since our usage of these methods is rather simple, for the sake of completeness, we will provide a short introduction here. We will restrict ourselves to the scalar-derived perturbations. Inclusion of the vector perturbations is rather immediate and does not change anything.

If $q_{ab}$ has a positive (negative) curvature, we may normalize it $q_{ab} = r_+^2 \mathring{q}_{ab}$ in such a way that $\mathring{R} \in \lbrace - 2,+2 \rbrace$. (We set  $\mathring{R} =0$ for a torus.) Let $\mathbb{S}$ be a non-constant eigenfunction of the Laplacian $\mathring{\Delta}$:
\begin{equation}
    \left( \mathring{\Delta} + k^2 \right) \mathbb{S} = 0
\end{equation}
and
\begin{subequations}
\begin{equation}
    \mathbb{S}_a = - \frac{1}{k} \mathring{D}_a \mathbb{S},
\end{equation}
\begin{equation}
    \mathbb{S}_{ab} = \frac{1}{k^2} \mathring{D}_a \mathring{D}_b \mathbb{S} + \frac{1}{2} q_{ab} \mathbb{S}.
    \label{eq:ttpiece}
\end{equation}
\end{subequations}
Then, we may decompose our perturbation as:
\begin{subequations}
\begin{equation}
    \delta F = f\,\mathbb{S}
\,,\quad
    \delta h_a = h\,\mathbb{S}_a
\,,\quad
    \delta q_{ab} = h_L\,\mathring{\gamma}_{ab} \mathbb{S} + h_T\,\mathbb{S}_{ab}
\end{equation}
\begin{equation}
    \delta E = q\,\mathbb{S}
\,,\quad
    \delta W_{a} = w\,\mathbb{S}_a
\,,\quad
    \delta Z_{a} = z\,\mathbb{S}_a
\,,\quad
    \delta B_{ab} = 0,
\end{equation}
\end{subequations}
where all new variables are simply constants. In this way, the problem of solving linearized Einstein--Maxwell reduces to solving a system of linear (algebraic) equations: 
\begin{subequations} \label{keyeqs}
\begin{equation}
    \left[\frac{k^2}{2} - \frac{1}{2}\,C\,r_{+}^2 \gamma\,(1+\gamma) - r_+^2 E^2 + \frac{3r_+^2}{L^2}\right] h_T - \frac{4\,E\,r_+^2 k}{\gamma} z +k^2 h_L = 0,
\end{equation}
\begin{equation}
    \frac{k^2-2\mathring{R}}{2k}\,E\,\gamma\,h_T + \left[C\,r_{+}^2\,(1+\gamma)\,\gamma -k^2- 4E^2 r_+^2\right] z + k\,E \left(\frac{3}{2}-\gamma \right) h_L = 0,
\end{equation}
\begin{equation}
    \gamma (\gamma-1) h_L = 0,
\end{equation}
\begin{equation}
    (\gamma+1)w + qk = 0,
\end{equation}
\begin{equation}
    q = \frac{k}{\gamma\,r_+^2}z - \frac{E}{r_+^2} h_L,
\end{equation}
\begin{equation}
    \frac{1}{2}\gamma(1+\gamma)\,h + \frac{k^2-\mathring{R}}{4k r_+^2}\gamma\,h_T - 2\,E\,z =0,
\end{equation}
\begin{equation}
        \frac{1}{2}(1+\gamma)(2+\gamma) f + \frac{1}{2r_+^2}(1+\gamma)\,k\,h - \frac{1}{r_+^2}\,C\,\gamma h_L +2\,E\,q = 0.
\end{equation}
\end{subequations}

Let us also mention that nonscalar derived perturbations would not introduce anything new. Indeed, there are no tensor perturbations in 4-dimensional spacetimes and an analogous calculation shows that the exponents for the vector perturbations are the same as for the scalar ones. 

Notice also that the case $\gamma=1$ is somewhat peculiar because then the system above is underdetermined. This is not a problem because there is also an additional gauge symmetry. All solutions with $\gamma=1$ on a sphere were found in \cite{Kolanowski:2021tje} -- they exist only for a fine-tuned values on $Q$. Our final formula (eq. \eqref{n_spherical} below) reproduces this result and so to simplify the discussion we will assume below that $\gamma\neq 1$ and consequently $h_L = 0$. 
This allows us to solve the first and the second equations which are decoupled from the rest. This is a system of two homogeneous equations for two unknowns and so it admits non-trivial solutions only when an appropriate determinant vanishes. Having found $h_T$ and $z$, we may solve all the other equations and every other variable is determined uniquely.
We will now go through the solutions in different cases, corresponding to the different topologies of $H$.

\subsubsection{Spherical black holes}
\begin{figure}[th]
\centering \includegraphics[width=0.7\textwidth]{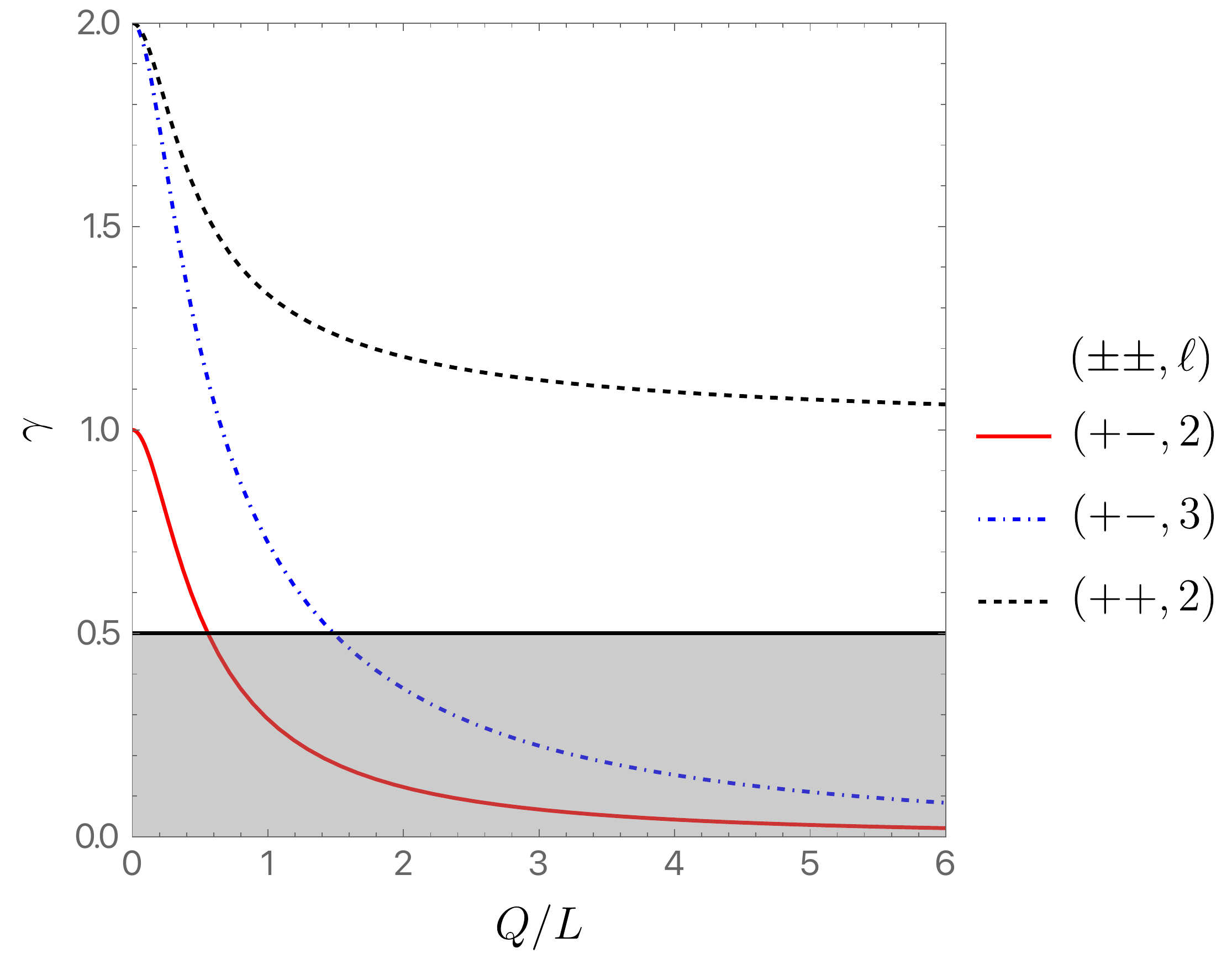}
\caption{\label{fig:spherical_RN_AdS}Scaling exponents for different $\ell$ as a function of $Q/L$. The shaded region indicates values of $\gamma$ for which no weak solution exits. The solid red line is a $+-$ mode with $\ell=2$,  the blue dash-dotted line is a $+-$ mode with $\ell=3$ and the black dashed line is a $++$ mode with $\ell=2$. All $\gamma < 2$ (except $\gamma =1$) lead to a singularity on the horizon via eq. \eqref{singweyl}. }
\end{figure}

We start by considering the case $H = S^2$. Then, $k^2 =  \ell (\ell+1)$, and $r_+$ is related to the electric charge $Q$ by \eqref{radiusQ}.
 Solutions to \eqref{keyeqs} exist only when
 \begin{subequations}
    \begin{equation}
    \gamma_{\pm \pm} = \frac{1}{2} \left[
    -1 \pm \sqrt{\frac{4\ell(\ell+1) + 5\sigma \pm 4 \sqrt{\sigma^2 + 2\ell(\ell+1)(1+\sigma)}}{\sigma}}
    \right], \label{n_spherical}
\end{equation}
where
\be{}\sigma \equiv 1 + 
\frac{6\,r_+^2}{L^2}.
\ee{}
\end{subequations}
There are a total of four solutions for each $\ell$ and $r_+$. We are free to choose a boundary condition at the horizon to remove two of them. If we choose combinations $-+$ and $--$, the exponent $\gamma$ is always negative so the solution blows up at the horizon. Thus, the physical perturbations are $++$ and $+-$. The values of $\gamma$ for a few $\ell$s are plotted in Fig. \ref{fig:spherical_RN_AdS}. As one may see, for $\ell =2,3$ and for $+-$ modes, the Weyl tensor is divergent for any value of $r_+$. Moreover, the larger $\frac{r_+}{L}$, the stronger the divergence. In particular, for sufficiently large black holes, the perturbation is not even a weak solution, and the region where no weak solutions exist is represented in gray. Since a generic nonspherical perturbation includes the $\ell=2$ mode,  we see that a generic  perturbation  would replace the horizon by a null singularity.

One may notice from Eq. \eqref{n_spherical} that $\gamma_{+-} = 0$ for $\ell =1$ (and any $r_+$). Since it does not decay, one might be tempted to interpret it as a deformation that changes the horizon geometry itself (and not just a neighborhood of it). This is however not justified since the system of equations \eqref{keyeqs} was derived using a decomposition into $\mathbb{S}$,$\mathbb{S}_a$ and $\mathbb{S}_{ab}$. When $\ell=1$, $\mathbb{S}_{ab} = 0$ and so all equations proportional to it are automatically satisfied. Instead, we are left with a simpler constraint ($k^2 = 2$):
\begin{equation}
\left[-2 + C r_{+}^2 (1+\gamma)\gamma - 4E^2 r_+^2\right] z = 0,
\end{equation}
which has non-trivial solutions only when
\begin{equation}
    \gamma_\pm = \frac{1}{2} \left(
    -1 \pm \sqrt{\frac{16+9\sigma}{\sigma}}
    \right).
\end{equation}   
Since we have only two solutions, our boundary conditions get rid of $\gamma_-$ (which would lead to the mode diverging at the horizon). Since $\gamma_{+}>1$, the solution is at least $C^1$ (but not necessarily $C^2$). 

For completeness, let us now discuss here what happens when $\Lambda \ge 0$. If $\Lambda = 0$, we have simply $\gamma_{+\pm} = \ell\pm1 \in \mathbb{N}$ and so perturbations are perfectly smooth. If $\Lambda >0$, $\gamma_{+ \pm}$ are generically not going to be integers so the perturbations have only finite degree of smoothness. In this case, the singularity is stronger for small black holes. In particular, if $\frac{r_+^2}{L^2}$ is small enough, it is only $C^1$ and it still suffers from the diverging tidal forces at the horizon. A static extremal black hole in our universe would certainly have small $\frac{r_+^2}{L^2}$, and nonspherical perturbations from other galaxies, so if they existed, they would have singular horizons.

\subsubsection{Toroidal black holes}

Analogous analysis can be performed in the toroidal case. This was done previously in \cite{Maeda:2011pk} for flat, non-compact cross-sections. As we will see, compactness changes the results qualitatively for small enough black holes. For definiteness we compactify the space directions in such a way that the cross-section of the horizon has volume $L_x\,L_y r_+^2$, where $x$ and $y$ are periodic coordinates with $x\sim x+L_x$ and $y\sim y+L_y$. From \eqref{background_parameters}, it follows that 
\begin{equation}
    E = \frac{\sqrt{3}}{L}
\end{equation}
and
\begin{equation}
    C =  \frac{6}{L^2}.
\end{equation}
Consequently, the charge confined within a black hole is:
\begin{equation}
    Q = \frac{\sqrt{3} L_x\,L_y r_+^2}{4\pi L}
\end{equation}
Note that in contrast to a sphere, we have a larger family of geometries on a two-dimensional torus -- it is encoded in periods of $x$ and $y$ coordinates and the angle between $\partial_x$ and $\partial_y$. 

We may now repeat our scheme and calculate the associated exponent. Perturbations are again scalar derived, however this time we need to decompose them into eigenfuctions of the Laplacian on $\mathbb{T}^2$, namely ${\rm Re}\,e^{i(k_x x + k_y y)}$ and ${\rm Im}\,e^{i(k_x x + k_y y)}$.
The exponent turns out to be\footnote{As before, there are four values of $\gamma$. Two of them are excluded automatically since they are negative. Below we focus only on the smaller positive value since we are interested  in possible singularities.} 
% \begin{equation}
%     \gamma = \frac{1}{2}\left(
%     -1 + \sqrt{1+13q^2 +3q^4 + \frac{15q^2}{\sqrt{1+q^2}} + \frac{3q^4}{\sqrt{1+q^2}}} 
%     \right),
% \end{equation}
% where
% \be{}
% q = \frac{k}{Q} \frac{L^3}{r_+^2} = \frac{k}{\sqrt{3} \pi} \frac{L^4}{r_+^4} .
% \ee
\begin{equation}
    \gamma = \frac{1}{6} \left(
    \sqrt{45 + 6\tilde{k}^2-36\sqrt{1+\frac{\tilde{k}^2}{3}}}-3
    \right),
\end{equation}
where $\tilde{k} \equiv k L/r_+$
with $k = \sqrt{k_x^2 + k_y^2}$.  As usual, the (linearized) Weyl tensor diverges when $\gamma<2$. Not surprisingly, for $k$ large enough, $\gamma$ given by the expression above is larger than $2$ and thus modes with high momenta are not singular, just as in the spherical case. However, if the black hole is toroidal, we can make it arbitrarily small {and then $\gamma$ is arbitrarily large.}
%(so that all non-zero values of $q$ are arbitrarily large). 
Thus, the toroidal black holes which are small enough are not going to become singular under an arbitrary (small) perturbation. It happens when 
% \begin{equation}
%     q^2 \gtrsim 0.844381,
% \end{equation}
\begin{equation}
    \tilde{k}^2 > 36 + 12 \sqrt{3}
\end{equation}
for all non-zero eigenvalues of the minus Laplacian $\mathring{\Delta}$. In particular, for a torus which is obtained from a square with $L_x = L_y = 2\pi$, the minimal $k$ is $1$ and this translates into
% \begin{equation}
%      0.44 L \gtrsim   r_+ .
% \end{equation}
\begin{equation}
    \frac{r_+}{L}<\frac{1}{\sqrt{36+12\sqrt{3}}}
\end{equation}
This explains the `almost all' in the title of this paper -- there is a finite volume in the moduli space of extremal black holes which are singularity-free. {On the opposite end of this spectrum are planar black holes for which there are infinitely many singular modes (of small momentum), as pointed out in \cite{Maeda:2011pk}}.

\subsubsection{Hyperbolic black holes}
Let us now consider a case when $H$ has a constant negative curvature. All such surfaces are given by a quotient of a hyperbolic space $\mathbb{H}^2$ by a discrete group. As follows from Eq. \eqref{background_parameters}, such solutions exist even without a Maxwell field. To begin, we will restrict ourselves to this case. As we noticed earlier, the larger the charge on a black hole, the more singular it becomes so we can expect that the inclusion of charge will make the perturbations even more divergent. This will be confirmed below. With $E=0$, it follows from \eqref{background_parameters} that $R = - \frac{6}{L^2}$ and so $r_+ = \frac{L}{\sqrt{3}}$. The analysis follows exactly in the same way as in the spherical case (except that only gravitational modes are included).

We find  that the perturbed solutions (for gravity modes) exist only when
\begin{equation}
    \gamma_{grav} = \frac{1}{2} \left(-1 
    + \sqrt{9+4k^2}
    \right).
\end{equation}
As before, the perturbation is singular if $\gamma_{grav}<2$, or equivalently $k^2 < 4$.  The first non-zero eigenvalue of the Laplacian on a compact Riemann surface is bounded by  \cite{yang1980eigenvalues, el1983volume}
\begin{equation}
    k^2 \le \frac{2}{g-1} \lfloor \frac{g+3}{2}\rfloor \le 4,
\end{equation}
where $g$ is a genus of $H$ and $\lfloor z\rfloor$ denotes the largest integer in $z$. Note that the last inequality is saturated only when $g=2$. We thus see that for $g>2$ and for any geometry of the horizon, at least one mode in the perturbation is singular. For $g=2$, a recent bootstrap calculation has shown that $k^2 \le 3.839$ \cite{Kravchuk:2021akc,Bonifacio:2021aqf}, so generic gravitational perturbations of all extremal hyperbolic black holes are singular. It is immediate to see that the same holds true also when the cross-section of the horizon is non-compact. Indeed, then the first non-zero eigenvalue is $\frac{1}{4}$ which is clearly less than $4$.

If we next consider a test Maxwell field on this neutral extremal black hole, the appropriate exponent reads
\begin{equation}
    \gamma_{EM} = \frac{1}{2} \left(-1 
    + \sqrt{1+4k^2}\right) < \gamma_{grav}.
\end{equation}
Notice that this perturbation may cause a singularity through its backreaction. Thus, the metric would be singular only if $\gamma_{EM} <1$ which translates to $k^2 < 2$. Thus, there are geometries on $H$ (for example, the Bolza surface which nearly saturates the above $g=2$ bound) for which the Maxwell field would not produce a singularity. Nevertheless, when the black hole is charged, the situation is very different.  Gravitational and Maxwell perturbations are then coupled to each other, and the two physical exponents become:
\begin{equation}
\gamma_{+\pm} = \frac{1}{2} \left[
-1 + \sqrt{
5 + \frac{4k^2 \pm 4 \sqrt{\sigma^2 + 2(\sigma-1)(k^2+2)}}{\sigma}
}
\right],
\end{equation}
where $\sigma = 6 \frac{r_+^2}{L^2} -1$. The minimal radius of the hyperbolic extremal horizon is obtained with no charge, $r_+ = \frac{L}{\sqrt{3}}$, so $\sigma \ge 1$. Notice that when $\sigma > \frac{1}{4} \left(4 + 2k^2 + k^4 \right)$, $\gamma_{+-}$ becomes {\it negative}. Thus the perturbation blows up on the horizon and our perturbative scheme breaks down. It is likely that some curvature invariants will now diverge. This also signals  an RG instability - a small change in the boundary conditions at asymptotic infinity (UV) would lead to a drastic change in the near horizon (IR) region. At the moment we are not sure what the endpoint of this instability is. Indeed, all smooth static near horizon geometries are classified in four dimensions \cite{Kunduri:2008tk} so the endpoint cannot be described by a single component extremal black hole with a smooth horizon. Most likely, the horizon just develops a strong singularity in which the metric is not even continuous.

\subsection{Kerr--AdS}
So far, we have considered only static solutions and their perturbation. This is of course far from any reasonable notion of `all extremal black holes'. To gain more completeness, let us now consider perturbations of the extremal Kerr-AdS with a mass $\frac{M}{(1-a^2/L^2)^2}$, an angular momentum $\frac{Ma}{(1-a^2/L^2)^2}$, and an angular velocity $\Omega$. It will be convenient to express these parameters in terms of the horizon radius $r_+\left(< \frac{L}{\sqrt{3}}\right)$:
\begin{equation}
    M = \frac{r_+ \left(1 + \frac{r_+^2}{L^2} \right)^2}{1 - \frac{r_+^2}{L^2}} \,,\quad a = r_+ \sqrt{ \frac{3r_+^2+L^2}{L^2 - r_+^2}} \,,\quad \Omega = \frac{\sqrt{L^4 + 2 r_+^2 L^2 - 3r_+^4}}{2r_+ L^2}
\end{equation}
Though very useful, $r_+$ has no geometric meaning \emph{per se}. Indeed, it is simply the location of the horizon measured in a particular coordinate system. We thus introduce the areal radius, defined as the square root of the area of the spatial section of the event horizon, divided by $4\pi$
\begin{equation}
R_+\equiv \frac{\sqrt{r_+^2+a^2}}{\sqrt{1-\frac{a^2}{L^2}}}\,.
\end{equation}

It is convenient to use the Teukolsky formalism \cite{Dias:2012pp}. This is especially useful since one works directly with the Weyl tensor. We want to consider only stationary perturbations. However, stationarity is ambiguous in this context since one could consider perturbations annihilated either by $\partial_t$ (which are stationary at infinity) or by the helical Killing vector $\partial_t + \Omega \partial_\phi$ (which are stationary at the horizon). Since we are interested in the behavior near the horizon, we choose the latter. Notice that this is the choice which corresponds to the black resonators at finite temperature \cite{Dias:2015rxy}. Following \cite{Dias:2012pp} we use the Kinnersly tetrad and Boyer-Lindquist coordinates for a non-rotating frame at infinity for the Kerr-AdS. Rather than restrict to pure gravitational perturbations, it is no more difficult to consider a spin $s$ perturbation (with $s={\pm}2$ corresponding to pure gravity). We want to separate variables for the spin $s$ field: 
\begin{equation}
    \Psi^{(s)} = e^{-i \omega t} e^{i m \phi} \Phi^{(s)}_{lm \omega}(r) S_{lm \omega} (\theta).
\end{equation}
Then, the radial (homogeneous) equation reads:
\begin{equation}
    \Delta_r^{-s} \partial_r \left[ 
    \Delta_r^{s+1} \partial_r \Phi^{(s)}(r) 
    \right] + H(r) \Phi^{(s)}(r) = 0 \label{teukolsky_radial}
\end{equation}
and the angular equation is
\begin{align}
\begin{split}
    \frac{1}{\sin \theta} \partial_\theta \left(
    \sin \theta \Delta_\theta \partial_\theta S_{lm \omega}
    \right) &+ \Bigg[ 
    (a \omega \cos \theta)^2 \frac{\Xi}{\Delta_\theta} - 2sa \omega \cos \theta \frac{\Xi}{\Delta_\theta} + s + \hat{\Lambda}_{\omega l m} 
     \\
    &- \left(m+s \cos \theta \frac{\Xi}{\Delta_\theta} \right)^2 \frac{\Delta_\theta}{\sin^2 \theta} - 2 \delta_s \frac{a^2}{L^2} \sin^2 \theta  \Bigg] S_{\omega l m}^{(s)} (\theta)= 0,
\end{split} \label{teukolsky_angular}
\end{align}
where
\begin{subequations}
\begin{align}
    \Delta_r &= (r^2 + a^2) \left(1 + \frac{r^2}{L^2} \right) - 2Mr \,,\quad
    \Xi = 1 - \frac{a^2}{L^2} \,,\quad
    \Delta_\theta = 1 - \frac{a^2}{L^2} \cos^2 \theta, \\
    \begin{split}
    H(r) &= \frac{K_T^2 - is \Delta_r' K_T}{\Delta_r} + 2 isK_T' + \frac{s+|s|}{2} \Delta_r'' - \hat{\lambda}_{\omega l m} \\
    &- |s|(|s|+1) (2|s| -1)(2|s|-7) \frac{r^2}{3L^2} - |s|(|s|-2)(4s^2 - 12|s|+11) \frac{a^2}{3L^2},
    \end{split} \\
    \begin{split}
        K_T &= \omega (r^2 + a^2) - ma \left(1 + \frac{r^2}{L^2} \right),
    \end{split} \\
    \hat{\lambda}_{\omega l m} &= \hat{\Lambda}_{\omega l m} - 2am\omega + a^2 \omega^2 + s + |s|,
\end{align}
\end{subequations}
$\hat{\Lambda}_{\omega l m}$ is a separation constant and $\delta_s = 1$ for $s \neq 0$ and $\delta_s=0$ for $s=0$.

\begin{figure}
        \centering
        \includegraphics[width=0.6\textwidth]{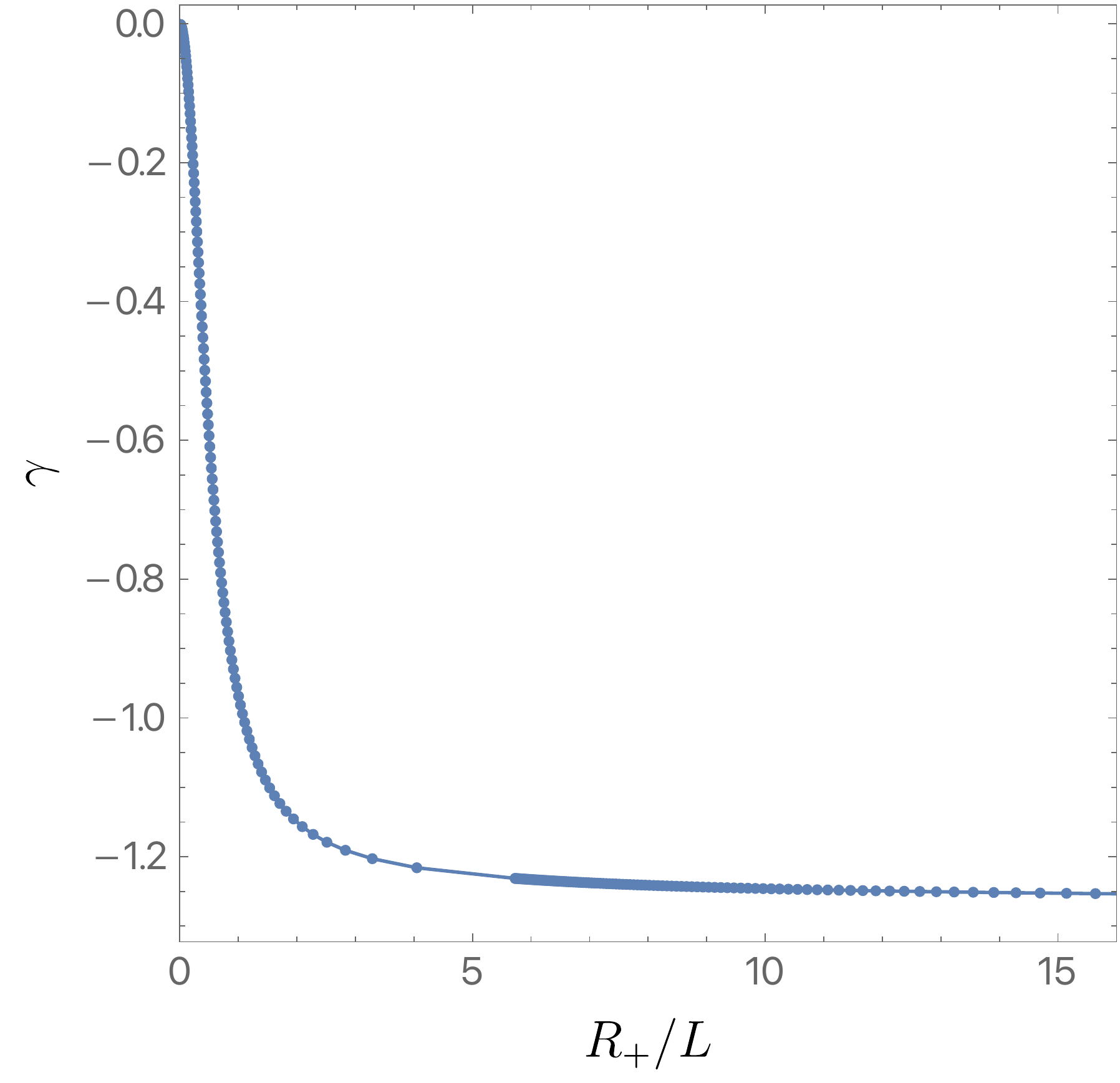}
        \caption{Curvature exponent $\gamma$ as a function of the area radius, $R_+/L$, for the $\ell=2, \ m=0$ mode of extremal Kerr-AdS. The fact that it is negative shows that the curvature diverges on the horizon.}
        \label{fig:Kerr-AdS}
    \end{figure}

Since we want the perturbation to be invariant under $\partial_t +\Omega \partial_\phi$, we only consider fields satisfying $\omega = \Omega m$.
Given $m \in \mathbb{Z}$, one first solves \eqref{teukolsky_angular} (treating it as an eigenproblem for $\hat{\Lambda}_{\omega l m}$) and then inserts the obtained values of that constant into the radial equation \eqref{teukolsky_radial}.  This
 equation for the extremal Kerr-AdS has a regular singular point at $r=r_+$. Thus, one finds that $\Psi_0 \sim (r-r_+)^\gamma$ and $\Psi_4 \sim (r-r_+)^{\gamma'}$.\footnote{Note that here $\gamma,\gamma'$ denote the exponents for the curvature perturbation, not the metric perturbation.} 
 As usual, there are two possible values of $\gamma$ and $\gamma'$ - one is always negative, so must be discarded. In general, $s=2$ modes (which describe the perturbation of $\Psi_0$) have smaller exponent. Their values for the $\ell =2, m=0$ mode are depicted in Fig. \ref{fig:Kerr-AdS}. 
 It is clear that this particular mode is always divergent. {Since $\Psi_0$ measures the tidal forces in the direction transversal to the horizon, this is the same type of singularity as the one encountered for static black holes.} 
 In general, increasing $\ell$ increases $\gamma$, so $\ell=3, m=0$ mode is non-singular for small black holes. If $m \neq0$, $\gamma$  {may become} {complex} ({even with $\Lambda = 0$}) -- this is a sign of the superradiant instability. 

For completeness, let us discuss what happens when one changes the cosmological constant. For simplicity, we restrict to $m=0$ perturbations.  In the Ricci flat $\Lambda = 0$ case, all the exponents are (with reasonable accuracy) integers. For $\Lambda > 0$, we found all $\gamma$ to be positive (at least while real). Thus, Kerr-(dS) seems to be free of this type of singularities at the horizon.  However, even a small charge appears to change this conclusion. In the Kerr-Newman dS background, the electromagnetic and gravitational perturbations are intertwined. On the neutral background, the former has a smaller scaling dimension $\gamma$. Since the introduction of a small charge will not change its value too much, it follows that this will cause diverging curvature for Kerr--Newman dS black holes. At the same time, when the cosmological constant vanishes, both scaling dimensions remain integer and thus no singularity develops.

%%%%%%%%%%%%%%%
\section{Einstein-Maxwell: nonlinear  results}
\label{sec_finite_temp}

In this section we show that the singularities we found in the linearized approximation actually occur in fully nonlinear solutions. In fact, they  leave a strong imprint on nonextremal black holes. In particular, we will see that although tidal forces are finite for $T > 0$, they become arbitrarily large as $T\to 0$.
 For simplicity, we will restrict our attention to static, topologically spherical black holes,  but the  results hold more generally. We start by explaining a scaling argument which allows us to relate powers of $\rho$ near an extremal horizon to powers of $T$ at a nonextremal horizon. We then explain how we numerically compute the nonlinear solutions. Finally, we present our results.
%%%%%%%%%%%%%%%
\subsection{\label{sec:scaling}Scaling argument}
%%%%%%%%%%%%%%%
Let us fix a charge $Q$ and a temperature $T \approx 0$. In  Schwarzschild coordinates, the horizon is now located at $r_h \approx r_0 + 2\pi T L_2^2$, where  
\begin{equation}
    L_2 = \frac{L r_0}{\sqrt{L^2 + 6r_0^2}}
\end{equation}
is the $AdS_2$ radius and $r_0$ is the radius of the horizon of the extremal black hole with the same charge. In the region $ \rho \equiv r-r_0 \ll r_0$, the metric can be approximated as
\begin{equation}
    g \approx \frac{\rho^2 - (r_h - r_0)^2}{L_2^2}{\rm d}t^2  + \frac{L_2^2}{\rho^2 - (r_h -r_0)^2} {\rm d}\rho^2 + (r_0 + \rho)^2 {\rm d}\Omega^2.
\end{equation}
The first two terms are just $AdS_2$ in Rindler coordinates. Thus, away from the horizon, any perturbation in this region  should behave in the same way as in $T=0$ case, i.e., the Weyl tensor should grow like
\begin{equation}
    C_{\rho a \rho b} \sim \rho^{\gamma-2},
\end{equation}
where $\gamma$ is given by \eqref{n_spherical}. When $\rho = O(r_h - r_0) = O(T)$, the Weyl tensor will be
\begin{equation}
    C_{\rho a \rho b}(\rho = O(T)) \sim T^{\gamma-2}.
\end{equation}
We do not know exactly what happens very closely to the horizon. Nevertheless, if the resulting spacetime is smooth, it cannot change very much. Thus, we
expect that on the horizon
\begin{equation}
    C_{\rho a \rho b}|_H \sim T^{\gamma-2}
\end{equation}
As we will now see, this is truly the case. This means, that even at finite $T$, tidal forces (though finite) can be arbitrarily large. 

It turns out that the singularities we discussed in the previous section have an important effect on standard black hole thermodynamics. Most notably, if $\gamma< 1/2$,  the increase in  black hole entropy with $T$ acquires an anomalous scaling at low $T$, which in turn yields an anomalous scaling for the specific heat at low temperatures. Intuitively, this can be seen as follows. From the previous section we know that the perturbation to the metric on the horizon decays like $\rho^\gamma$, but the linearized analysis  shows that there is no change to the volume element to leading order.   So the leading correction to the area comes from a second order contribution which scales like $\rho^{2\gamma}$. We now use the above scaling argument to relate $\rho$ to the near extremal horizon temperature $T$ to obtain
\begin{equation}
S \approx S_0+S_2\,T^{2\gamma}
\end{equation}
where $S_0$ and $S_2$ are suitable constants. This implies that the specific heat at constant charge scales like 
\be\label{anomcq}
C_Q = T \frac{{\rm d}S}{{\rm d}T}\ \propto\ T^{2\gamma}
\ee

We  confirm this scaling for Einstein-Maxwell theory using second order perturbation theory about the near horizon geometry of an extreme RN-AdS black hole in the Appendix. We also show  that for $\gamma = 1/2$, there is an anomalous $T \log T$ scaling of the specific heat. 

We have discussed so far how the perturbations behave near the (nearly extremal) horizon but we omitted their source. The easiest way to obtain such solutions is to consider slightly deformed boundary conditions. Indeed, a standard (spherically symmetric) black hole spacetimes satisfy
\begin{equation}
    A_t|_{\partial M} = \mu_0,
\end{equation}
where $\mu_0$ is a constant. By the AdS/CFT dictionary, it corresponds to the chemical potential for the dual theory. We perturb this boundary condition by writing
\begin{equation}
    A_t|_{\partial M} = \mu(\theta,\phi), \label{BC_background}
\end{equation}
where $\mu$ is still time-independent. We look for static black hole solutions (of fixed temperature $T$) satisfying this condition and we monitor their Weyl tensor at the horizon.\footnote{We will see the anomalous scaling of the specific heat in the next section, in a theory where it is easier to reach the low temperatures required.} Note that in this scheme, the total charge $Q$ is determined by the solution and not prescribed apriori. Of course, we also need  to specify the metric at infinity. For simplicity we will keep it spherical.  

A few words about $\mu(\theta,\phi)$ are in place. Since it is a function on a sphere, we may decompose it into spherical harmonics
\begin{equation}
    \mu(\theta,\phi) = \sum_{\ell, m} \mu_{\ell, m} Y_{\ell m}(\theta,\phi).
\end{equation}
Although we are interested in the full, non-linear theory, let us quickly present what would happen in a perturbative scheme. If we treat a deviation from  spherical symmetry as a small correction, non-spherically symmetric contributions satisfy linearized Einstein-Maxwell equations on a Reissner--Nordstr\"om-AdS background with a boundary condition \eqref{BC_background}. Then, modes with different $\ell, m$ decouple. From \eqref{n_spherical}, we can see that $\ell=2$ would  dominate near the horizon. However, the theory in question is highly non-linear and we may reasonably expect that if we turn on one mode at infinity, all are going to be non-zero near the horizon. Thus, for generic $\mu(\theta,\phi)$ we should obtain the exponent $\gamma$ given by \eqref{n_spherical} with $\ell=2$. Nevertheless, if $\mu_{2,m}$ is much smaller than different components, it may happen that the amplitude in front of
\begin{equation}
    C_{\rho a \rho b} \sim T^{\gamma-2}
    \label{eq:scalingW}
\end{equation}
is going to be very small and this universal effect would become visible only at very low temperatures.

%%%%%%%%%%%%%%%%%%%%%%%%%%%%%%%%%%%%
\subsection{Numerical scheme\label{sec:num}}
%%%%%%%%%%%%%%%%%%%%%%%%%%%%%%%%%%%%
%The aim of this section is to show that the linear results of Sec. 3 are valid at the full nonlinear level. In order to do this, we will change slightly gears and work at finite temperature, in the hope of checking the scaling of Eq.~(\ref{eq:scalingW}).

To construct solutions at nonzero $T$, we will work directly in Bondi-Sachs coordinates.  The reason for this is two-fold: 1) in Bondi-Sachs coordinates there are no non-analyticities near the conformal boundary, and we thus expect numerical methods to exhibit stronger convergence properties; 2) we need to work in a coordinate system where ingoing null geodesics are easy to determine, so that one can easily calculate tidal force singularities. Our work will roughly follow \cite{Chesler:2013qla,Biggs:2022lvi}, but with some important differences.

We thus take the following metric and gauge field Ans\"atze 
\begin{subequations}
\begin{multline}
\mathrm{d}s^2=\frac{L^2}{y^2}\Bigg\{-(1-y)(1+y \,A_1)\frac{\mathrm{d}v^2}{L^2}-\frac{2\,\mathrm{d}v\,\mathrm{d}y}{L}
\\
+A_2\left[A_3\left(\mathrm{d}x+\frac{A_4(1-y)\sin x\ \mathrm{d}v}{L}\right)^2+\frac{\sin^2 x}{A_3}\mathrm{d}\phi^2\right]\Bigg\}\,,
\end{multline}
\label{eqs:ansatze}%
\begin{equation}
A=(1-y)A_5\mathrm{d}v+L\,\sin x\,A_6\,\mathrm{d}x\,,
\end{equation}
\end{subequations}%
where all $A_i$ are functions of $(x,y)$ and $\phi$ is a periodic coordinate with $\phi\sim\phi+2\pi$. Note that $x\in[0,\pi]$ is an angular coordinate and $y=[0,1]$ is a radial coordinate, with $x=0,\pi$ being the poles of the two-sphere and $y=0,1$ being the conformal boundary and black hole event horizon, respectively. We note that $\partial/\partial y$ is \emph{globally} null and is thus a physical coordinate with respect to which one can easily compute tidal force singularities by looking at certain components of the Weyl tensor in $(v,y,x,\phi)$ coordinates.

It remains to explain how Eqs.~(\ref{eqs:ansatze}) together with the Einstein-Maxwell equations give rise to a well defined system of Elliptic equations which can be solved using standard numerical methods, such as the ones in \cite{Dias:2015nua}.

Instead of using the trace-reversed version of the Einstein equation, we are going to use the Einstein equation itself. The reason for this is that we will take advantage of the constraint equations to proceed. We thus consider the following equations of motion 
\begin{subequations}
\begin{equation}
E_{\mu\nu}\equiv R_{\mu\nu}-\frac{R}{2}g_{\mu\nu}-\frac{3}{L^2} g_{\mu\nu}-2\left(F_{\mu\alpha}F_{\nu}^{\phantom{\mu}\alpha}-\frac{g_{\mu\nu}}{4}F_{\alpha\beta}F^{\alpha\beta}\right)=0\,,
\end{equation}
\begin{equation}
P_\mu\equiv \nabla^\nu F_{\nu\mu}=0\,.
\end{equation}
\end{subequations}%
Note that we have more independent components of the Einstein equations than variables to solve for. We thus need to choose a subset of the equations and show that, given appropriate boundary conditions, the remaining equations are solved. We thus label the equations we actually solve for as dynamical, and the ones that are enforced via the Bianchi identities as constraints (in analogy with the constraint and dynamical equations of the initial value problem in general relativity).

For dynamical equations we take $E^{\mu\nu}$ and $P^{\mu}$ with $\mu=\{v,x,\phi\}$. This gives us exactly six equations to solve for within our symmetry class. For the constraint equations we take $E^{y\mu}$ and $P^y$. Now, we would like to show that if we impose $E^{y\mu}$ and $P^y$ at either $y=0$ or $y=1$, then by virtue of the dynamical equations and the Bianchi identities, the constraint equations $E^{y\mu}$ and $P^y$ should be satisfied everywhere\footnote{It might appear confusing that we only need to impose the constraint at a single boundary, but this is the result of the Bianchi identities being first order in $E_{\mu\nu}$.}. This is a relatively simple exercise that we leave to the reader.  We do, however, have to check that the dynamical equations form a well posed Elliptic problem. This, in particular, means that near each boundary we must have a number of free constants of integration that matches the order of the differential equation. That is to say, near the conformal boundary and black hole event horizon we expect to find six functions of integration.

In four spacetime dimensions, when using Bondi-Sachs coordinates and focusing on the Einstein-Maxwell equations, one can show  that the asymptotic expansion of the $A_i$ functions near the boundary is power law in $y$. As such, we take
\begin{equation}
A_i(x,y) = \sum_{I=0}^{+\infty} y^{I}A_i^{(I)}(x)\,.
\end{equation}
As boundary conditions we impose
\begin{equation}
A_2(x,0)=A_3(x,0)=1\,,\quad A_4(x,0)=A_6(x,0)=0\,,\quad\text{and}\quad A_5(x,0)=\mu(x)\,,
\end{equation}
with $\mu(x)$ being our boundary chemical potential. Note that we have not yet detailed what boundary condition we take for $A_1$. One can input the above expansion in the dynamical equations and work out which coefficients are left free. It turns out that all remaining coefficients are uniquely fixed in terms of $\mu(x)$ and
\begin{equation}
\left\{A_1^{(0)}(x),A_1^{(1)}(x),A_1^{(2)}(x),A_3^{(3)}(x),A_4^{(3)}(x),A_5^{(1)}(x),A_6^{(1)}(x)\right\}
\end{equation}
and their derivatives with respect to $x$. These are seven free functions (once we fix $\mu(x)$), and thus one too many with respect to what we are expecting. To remove this redundancy we impose the constraint equation $E^{yv}$ asymptotically. This constraint then imposes
\begin{equation}
\left.\frac{\partial A_1}{\partial y}\right|_{y=0}=\frac{1}{4} \left[5+2 A_1(x,0)+A_1(x,0)^2\right]\,,
\end{equation}
which fixes $A_1^{(1)}(x)$ in terms of $A_1^{(0)}(x)$. We thus have the expected number of integration functions near the conformal boundary, and have imposed one of the constraint equations.

We now turn our attention to the black hole event horizon located at $y=1$. The powers of $(1-y)$ in the Ansatz for the gauge field and metric (see Eqs.~(\ref{eqs:ansatze})) were chosen in such a way that $P^y$ and $E^{yx}$ are automatically satisfied at the black hole event horizon. The savy reader will note that we have not yet imposed $E^{yy}$, but we shall shortly. We again expand all equations in power series in $(1-y)$ near the black hole event horizon
\begin{equation}
A_i(x,y) = \sum_{I=0}^{+\infty} (1-y)^{I}\tilde{A}_i^{(I)}(x)\,.
\label{eq:hor}
\end{equation}
We also impose $E^{yy}$ at the horizon, which demands that $\tilde{A}_1^{(0)}$ be a constant. In fact, if we set
\begin{equation}
\tilde{A}_1^{(0)}=T_0-1
\end{equation}
we find that the black hole temperature is given by
\begin{equation}
4\pi L T=T_0\,.
\end{equation}
All coefficients appearing in the expansion near the horizon Eq.~(\ref{eq:hor}) are uniquely fixed in terms of $T_0$ and
\begin{equation}
\left\{\tilde{A}_1^{(1)}(x),\tilde{A}_2^{(0)}(x),\tilde{A}_3^{(0)}(x),\tilde{A}_4^{(0)}(x),\tilde{A}_5^{(0)}(x),\tilde{A}_6^{(0)}(x)\right\}\,,
\end{equation}
which is precisely the number of expected integration constants near $y=1$.

All we need to discuss are the axes of symmetry, located at $x=0,\pi$. We will focus on $x=0$, but $x=\pi$ has identical boundary conditions. Since we want $\phi$ to have period $2\pi$, we must demand $A_3(0,y)=A_3(\pi,y)=1$. Regularity at the axis further imposes
\begin{equation}
\left.\frac{\partial A_i}{\partial x}\right|_{x=0}=0\,.
\end{equation}
One can check that the above boundary conditions are consistent with the constraint equations, and provide the right number of integration functions near the axis of symmetry. We have thus sketched in some detail that our dynamical equations, together with our choice of boundary conditions, enforces the constraint equations and gives rise to a well posed Elliptic problem.

Before presenting the results, we shall detail a little the numerical method we used. Since the functions $A_i$ develop enormous gradients close to the black hole event horizon as $T\to0$, we discretise the integration domain into two patches $[0,y_c]\cup[y_c,1]$. Chebyshev-Gauss-Lobatto grids are placed in each patch using transfinite interpolation (see \cite{Dias:2015nua}). At the patch boundaries we require the metric and its first derivative to be continuous (continuity of the remaining derivatives is then enforced via the equations of motion). We then implement a standard Newton-Raphson routine, with each linear iteration being solved via a LU decomposition. Since large gradients develop near the horizon, we typically take $y_c=0.95$. On the patch near the event horizon we had to resort to enormous resolutions (with more than $100$ points in the $y$ direction) to resolve the gradients and achieve convergence. With these high resolutions we were able to reach as low as $4\pi L T =10^{-3}$.
%%%%%%%%%%%%%%%
\subsection{Results}
%%%%%%%%%%%%%%%
For the boundary profile we take
\begin{equation}
\mu(x)=\bar{\mu}+\mu_1\,\cos x\,.
\end{equation}
This corresponds to a dipole type perturbation, where the inhomogeneous component of the chemical potential is given entirely in terms of an $\ell=1$ harmonic on the round two-sphere.

\begin{figure}[th]
\centering \includegraphics[width=0.6\textwidth]{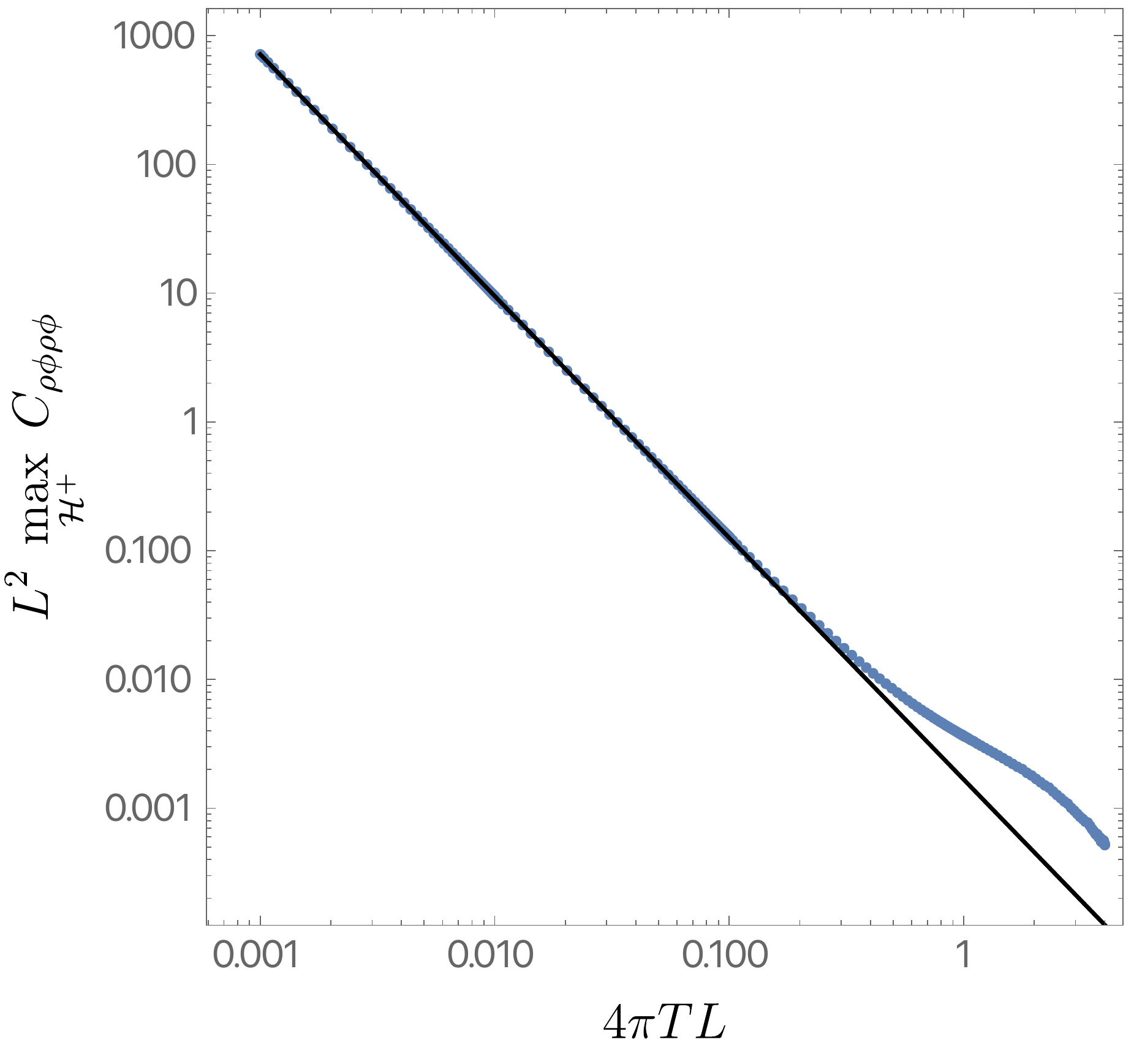}
\caption{\label{fig:scalingWmaxwell} A plot showing the maximum of $ C_{\rho\phi\rho\phi}$ on the horizon as a function of $4 \pi T L$, computed for $\bar{\mu}=2$ and $\mu_1=0.1$. The solid black line shows the best fit to the functional form given in Eq.~(\ref{eq:fitW}) in the range $4 \pi T\in[10^{-3},10^{-2}]$ yielding $\tilde{\gamma}\approx 0.122401$ and $a_0\approx 0.00166763$. The blue disks correspond to the numerical data points extracted from our simulations.}
\end{figure}

The most interesting quantity to plot is the $C_{\rho\phi\rho\phi}$ component of the Weyl tensor evaluated on the event horizon, \emph{i.e.} at $y=1$, for which the scaling given in Eq.~(\ref{eq:scalingW}) should apply. In Bondi-Sachs coordinates, this component is simply given by
\begin{equation}
C_{\rho\phi\rho\phi}=-\frac{\sin ^2x}{2 y^2 A_3^2} \left[\frac{A_2}{A_3}\left(\frac{\partial A_3}{\partial y}\right)^2-\frac{\partial A_2}{\partial y} \frac{\partial A_3}{\partial y}-A_2 \frac{\partial^2 A_3}{\partial y^2}\right]\,.
\end{equation}
In Fig.~\ref{fig:scalingWmaxwell} we plot the maximum value of $C_{\rho\phi\rho\phi}$ on the black hole event horizon, \emph{i.e.} $\max_{\mathcal{H}^+}C_{\rho\phi\rho\phi}$, as a function of $4\pi T L$ in a log-log scale, for fixed values of $\bar{\mu}=2$ and $\mu_1=0.1$. We have tried other values of $\bar{\mu}$ and $\mu_1$ and the results remain qualitatively similar. If the scaling given in Eq.~(\ref{eq:scalingW}) holds, we should see a straight line in a log-log plot appearing at low $T$. This appears evident from Fig.~\ref{fig:scalingWmaxwell}. Furthermore, the slope of this linear behaviour should be given precisely by $\gamma-2$, with $\gamma$ being computed by the limiting $Q$ at zero temperature. We can extract such $Q$ from the numerical results via a linear fit
\begin{equation}
\frac{G_4 Q}{L} = q_0+q_1 T\,
\end{equation}
in the range $4 \pi TL\in[10^{-3},10^{-2}]$. We find $q_0\approx 2.00181$ which, as expected is very close to $\bar{\mu}$, since $\mu_1$ is small.  This is the value of $Q$ that that controls the smallest value of $\gamma$. This occurs for the $\ell=2$ mode with the minus sign in Eq.~(\ref{n_spherical}). For this value of $Q$, we find $\gamma \approx0.122025$. On the other hand, the solid black line in Fig.~\ref{fig:scalingWmaxwell} is well fit by a function of the form
\begin{equation}
a_0 T^{\tilde{\gamma}-2}
\label{eq:fitW}
\end{equation}
and via a fit in the region $4 \pi T\in[10^{-3},10^{-2}]$ we find $\tilde{\gamma}\approx0.122401$ and $a_0\approx 0.00166763$. The agreement between $\gamma$ computed with the limiting value of $Q$ and according to Eq.~(\ref{n_spherical}) and the value extracted from the fit ($\tilde{\gamma}$) is striking: the difference is below $0.5\%$. Furthermore, we expect the $\ell=2$ mode to be nonlinearly sourced by the $\ell=1$ boundary mode. As such, the expectation is that this mode at the horizon must have a magnitude that scales as a power of $\mu_1$.  This is precisely what we find for $a_0$.

\begin{figure}[th]
\centering \includegraphics[width=0.6\textwidth]{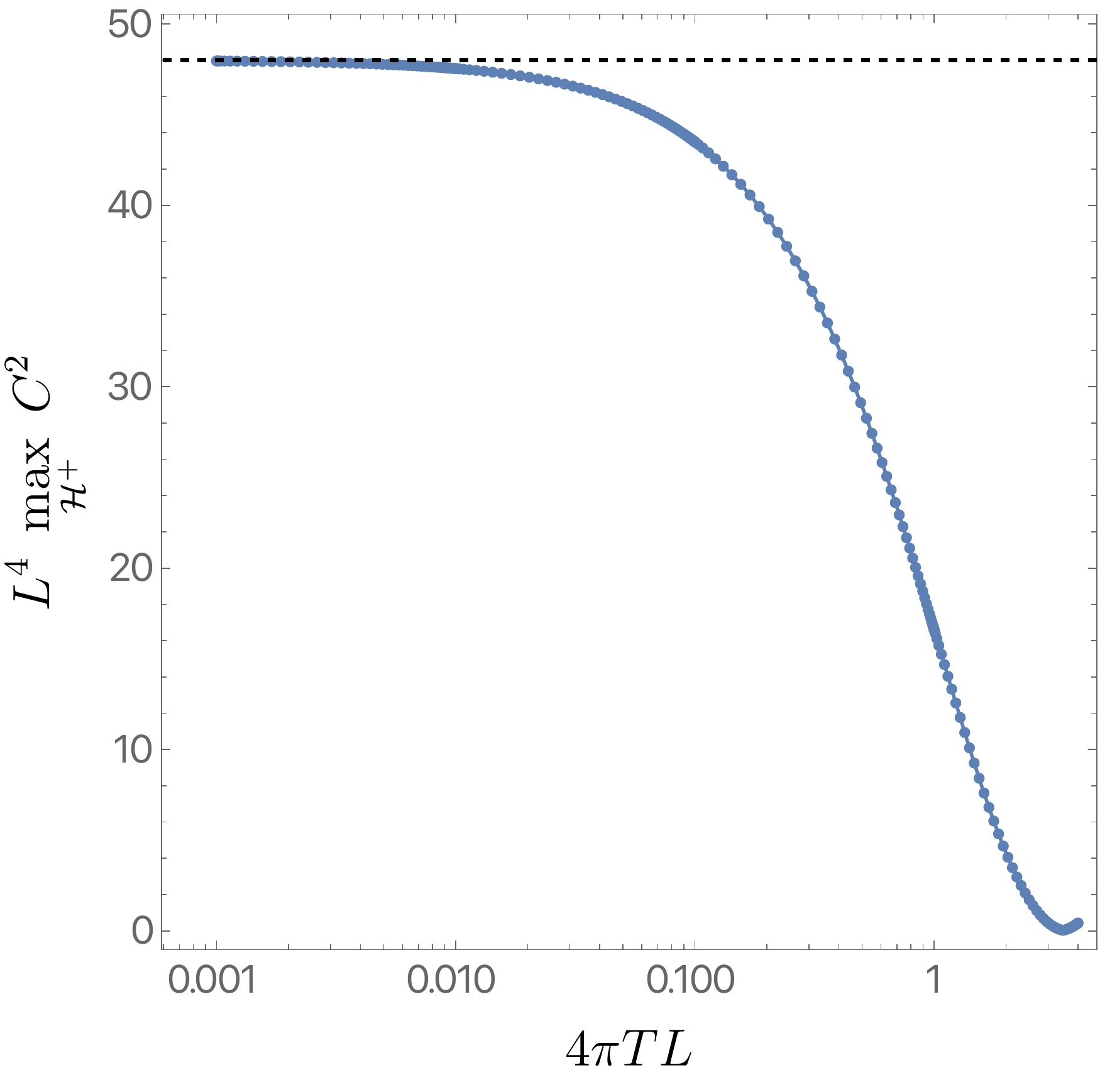}
\caption{\label{fig:maximumwel} A plot showing the square of the Weyl tensor on the black hole event horizon as a function of $4 \pi T L$, computed for the same black holes as Fig. \ref{fig:scalingWmaxwell}. The solid black line shows the $AdS_2$ prediction of Eq.~(\ref{eq:preads2}). The blue disks correspond to the numerical data points extracted from our simulations.}
\end{figure}

We have also looked at curvature invariants, such as the maximum value of the \emph{square} of the Weyl tensor at the black hole horizon. For an extremal RN-AdS black hole of \emph{any} size we find that the square of the Weyl tensor on the horizon $\mathcal{H}^+$ is simply
\begin{equation}
\left.L^4 C^2\right|_{\mathcal{H}^+}\equiv L^4 \left.C^{\mu\nu\rho\lambda}C_{\mu\nu\rho\lambda}\right|_{\mathcal{H}^+}=48\,.
\label{eq:preads2}
\end{equation}
If our linear results are to hold, we should see all curvature invariants approaching their $AdS_2$ value at small temperatures. In Fig.~\ref{fig:maximumwel} we plot the maximum value of the square of the Weyl tensor on the black hole event horizon as a function of the temperature in a $\log$-linear plot. The blue disks are the exact numerical results and the dashed black horizontal line is the $AdS_2$ prediction. Clearly, the square of the Weyl tensor on the horizon not only remains finite as we cool down the system but it also approaches its unperturbed  value, as predicted by perturbation theory.

%%%%%%%%%%%%%%%%%%%%%%%
\section{Nonlinear scalar field model}
%%%%%%%%%%%%%%%

It is difficult to numerically construct the $T=0$ limit of the solution in the previous section, or get $T$ low enough to see the anomalous scaling of the specific heat predicted in Sec. 4.1. To remedy this, in this section we consider a simpler model in which both can be achieved. We  confirm that the exact $T=0$ solution has diverging tidal forces as predicted by the linear analysis, and that the specific heat has anomalous scaling.

 Our model consists of adding $2\ell+1$ neutral, massless scalar fields to our Einstein-Maxwell theory. By cleverly choosing their angular dependence, one can keep the stress tensor spherically symmetric, so the metric remains spherical. Nevertheless, from the behavior of a test field in Sec. \ref{sec:simpe}, we expect the scalars will develop powerlaw behavior with noninteger exponents near the horizon which will backreact on the metric. 

The scalar fields can be viewed as components of a vector $\vec{\Phi}$:
\begin{equation}
(\vec{\Phi})_m=\phi_m\,,
\end{equation}
with $|m|\leq \ell$.
The action is then simply
\begin{equation}
S=\frac{1}{16\pi G}\int_{\mathcal{M}}\mathrm{d}^4x\,\sqrt{-g}\left[R+\frac{6}{L^2}-F_{\mu\nu}F^{\mu\nu}-(\nabla_\mu \vec{\Phi})\cdot (\nabla^\mu \vec{\Phi})\right]\,.
\label{eq:simp}
\end{equation}
where $\cdot$ denotes the usual Cartesian dot product between vectors in $\mathbb{R}^{2\ell+1}$ and $G$ is Newton's constant. One could generalize this by adding the same mass to all scalars without changing the conclusion.

The equations of motion derived from (\ref{eq:simp}) read
\begin{subequations}
\begin{equation}
R_{\mu\nu}-\frac{R}{2}g_{\mu\nu}-\frac{3}{L^2}g_{\mu\nu}=2\left(F_{\mu}^{\phantom{a}\alpha}F_{\nu\alpha}-\frac{g_{\mu\nu}}{4}F_{\alpha\beta}F^{\alpha\beta}\right)+(\nabla_\mu \vec{\Phi})\cdot(\nabla_\nu \vec{\Phi})-\frac{g_{\mu\nu}}{2}(\nabla_\alpha \vec{\Phi})\cdot(\nabla^\alpha \vec{\Phi})\,,
\end{equation}
\begin{equation}
\nabla^\nu F_{\mu\nu}=0\,,
\label{eq:maxwellsimp}
\end{equation}
and
\begin{equation}
\nabla_\mu \nabla^\mu \vec{\Phi}=0\,.
\end{equation}
\end{subequations}

We will focus on static black hole solutions whose stress energy tensor is spherically symmetric. We then introduce ingoing Eddington-Finkelstein coordinates $(v,r,\theta,\phi)$ in which the metric and gauge field take the form
\begin{subequations}
\begin{equation}
\mathrm{d}s^2=-f(r)\,\mathrm{d}v^2+2\,\mathrm{d}v\,\mathrm{d}r+r^2\,p(r)\left(\mathrm{d}\theta^2+\sin^2\theta \mathrm{d}\phi^2\right)\,,
\label{eq:linesimp}
\end{equation}
and
\begin{equation}
A=A_v(r)\,\mathrm{d}v
\end{equation}
\end{subequations}%
with  $f$ and $p$ functions of $r$ to be determined by our numerical scheme.

As promised, we would like to retain non-homogeneity in the scalar field, and at the same time ensure that the resulting stress energy momentum tensor is spherically symmetric. We can achieve this by choosing
\begin{equation}
\phi_m(\hat{r},\theta,\phi) = \phi(r)\times \sqrt{\frac{(\ell-m)!}{(\ell+m)!}}\times P^{\ell}_m(\cos \theta)\times\left\{
\begin{array}{ll}
\sqrt{2}\,\sin (m\,\phi)\,,& \text{for}\quad m<0
\\
1\,, & \text{for}\quad m=0
\\
\sqrt{2}\,\cos (m\,\phi)\,,& \text{for}\quad m>0
\end{array}\right.\,,
\end{equation}
 where $P^{\ell}_m(x)$ are the standard associated Legendre polynomials. The normalisations are chosen so that 
 \begin{equation}
 \sum_{m=-\ell}^{\ell} \phi_m(r,\theta,\phi)^2=\phi(r)^2\,.
 \end{equation}

 We first solve Eq.~(\ref{eq:maxwellsimp}). This is a simple exercise, since the scalars are uncharged under the Maxwell field and the solution is spherically symmetric. We thus find
 \begin{equation}
 A_v^\prime(r)=\frac{Q}{r^2\,p(r)}\,,
 \end{equation}
 where $Q$ is an integration constant and $'$ denotes differentiation with respect to $r$. It is a simple exercise to show that in fact $Q$ is the charge of the black hole solutions we seek to construct so long as we demand $\lim_{r\to+\infty}p(r)=1$. We can use this relation in the remaining Einstein and scalar field equations to eleminate all dependence on $A_v$. The Einstein and scalar field equations reduce to
 \begin{subequations}
 \begin{align}
\frac{(r^2\,p)'}{p}\frac{f'}{f}+\frac{1}{2}\frac{(r^2p)'^2}{r^2p^2}-\frac{6 r^2}{fL^2}+\frac{2 Q^2}{r^2\,f\,p^2}-\frac{2}{f\,p}+\frac{\ell  (\ell +1) \phi^2}{f\,p}-r^2 \phi'^2&=0\,,\label{eq:fsimp}
\\
\frac{1}{r^2p}(r^2\,p\,f\,\phi')'-\frac{\ell(\ell+1)}{r^2p}\phi&=0\,,\label{eq:phisimp}
\\
\frac{1}{r^2p}(r^2p')'+\phi'^2-\frac{p'^2}{2p^2}&=0\,.
 \end{align}
 \label{eqs:simp}
 \end{subequations}%%
 
 The line element Eq.~(\ref{eq:linesimp}) has a residual gauge freedom. Namely, we are free to shift $r$ by a constant. We fix this freedom by demanding $p(r_+)=1$, where $r=r_+$ is the black hole event horizon (extremal or otherwise) defined by $f(r_+)=0$.

 %%%%%%%%%%%%%%%%%%%%%%%%%%%%%%%%%%%%%%%%%%
 \subsection{Zero temperature results}
 %%%%%%%%%%%%%%%%%%%%%%%%%%%%%%%%%%%%%%%%%%
 For zero temperature black holes, we want $f(r)$ to have a double zero at the black hole horizon $r=r_+$.  We shall also assume that near the horizon $\phi\propto(r-r_+)^\gamma$ for some power $\gamma>0$ to be determined shortly.
 From Eq.~(\ref{eq:fsimp}) it easy to see that the requirement that $f$ has a double zero at $r=r_+$ implies
 \begin{equation}
 Q=r_+\,\sqrt{1+\frac{3\,r_+^2}{L^2}}\,.
 \end{equation}
 We thus choose this value throughout and parameterise our extremal black holes by $r_+/L>0$.
 
 We now write
 \begin{equation}
 f = f_0(r-r_+)^2\left(1+\delta f\right)\,,
 \end{equation}
 where $\delta f$ is a function we wish to determine next as a function of $\gamma$. The first nontrivial order in Eq.~(\ref{eq:fsimp}) determines $f_0$ as a function of $r_+$, and we  find
 \begin{equation}
 f_0=\frac{L^2+6 r_+^2}{L^2 r_+^2} = \frac{1}{L_2^2}\,.
 \end{equation}
 where $L_2$ is the $AdS_2$ radius. The next to leading order terms in Eq.~(\ref{eq:fsimp}) determine $\delta f$. The leading contribution to $\delta f$ is of the form
 \begin{equation}
 (r-r_+)^2\delta f^\prime \ \propto\ (r-r_+)^{2\hat{\gamma}+1}\Rightarrow \delta f \ \propto\ (r-r_+)^{2\hat{\gamma}}\,,
 \end{equation}
 where $\hat{\gamma}=\min\;(\gamma,1/2)$.
 
Similarly, $p$ admits an expansion 
 \begin{equation}
 p=1+p_1 (r-r_+)^{2\hat{\gamma}}\,,\label{eq:fp}
 \end{equation}
 where $p_1$ is a known constant.
 For $\gamma=1/2$, there are logarithmic terms appearing in the expansions of both $f$ and $p$.
 
 It remains to determine $\gamma$. These we read off from the scalar equation, which to leading order off the extremal horizon yield 
 \begin{equation}
 \gamma=\frac{1}{2}\left[\sqrt{1+\frac{4\ell(\ell+1)}{1+6\,y_+^2}}-1\right]\,,
 \label{eq:n}
 \end{equation}
 where $y_+$ is again the dimensionless radius $y_+\equiv r_+/L$.  Note that this agrees with the linear result \eqref{eq:n_KG}.
 
To proceed numerically, we introduce a compact coordinate
 \begin{equation}
 r=\frac{r_+}{1-y}\,
 \label{eq:compact}
 \end{equation}
 so that the extremal horizon is located at $y=0$, whereas the conformal boundary is located at $y=1$. To implement our boundary conditions, we also define 
 \begin{equation}
 f(r) \equiv \frac{\left(r-r_+\right)^2 \left(L^2+r^2+3 r_+^2+2 r r_+\right)}{L^2 r^2}q(r)\,.
 \label{eq:defg}
 \end{equation}
 and regard $q$, $\phi$ and $p$ as functions of $y$. The boundary conditions are now simply $q(0)=1$, $\phi(0)=0$, $p(0)=1$ at the extremal horizon. At the conformal boundary we demand $q(1)=1$, $\phi(1)=V$ and $p(1)=1$. The conditions on $q$ and $p$ just ensure that the spacetime is asymptotically AdS.  $V$ is the amount that we turn on the scalars asymptotically. If there were a holographic dual to this theory,  $V$ would be the source of the operator dual to $\phi$.
 
\begin{figure}[th]
\centering \includegraphics[width=\textwidth]{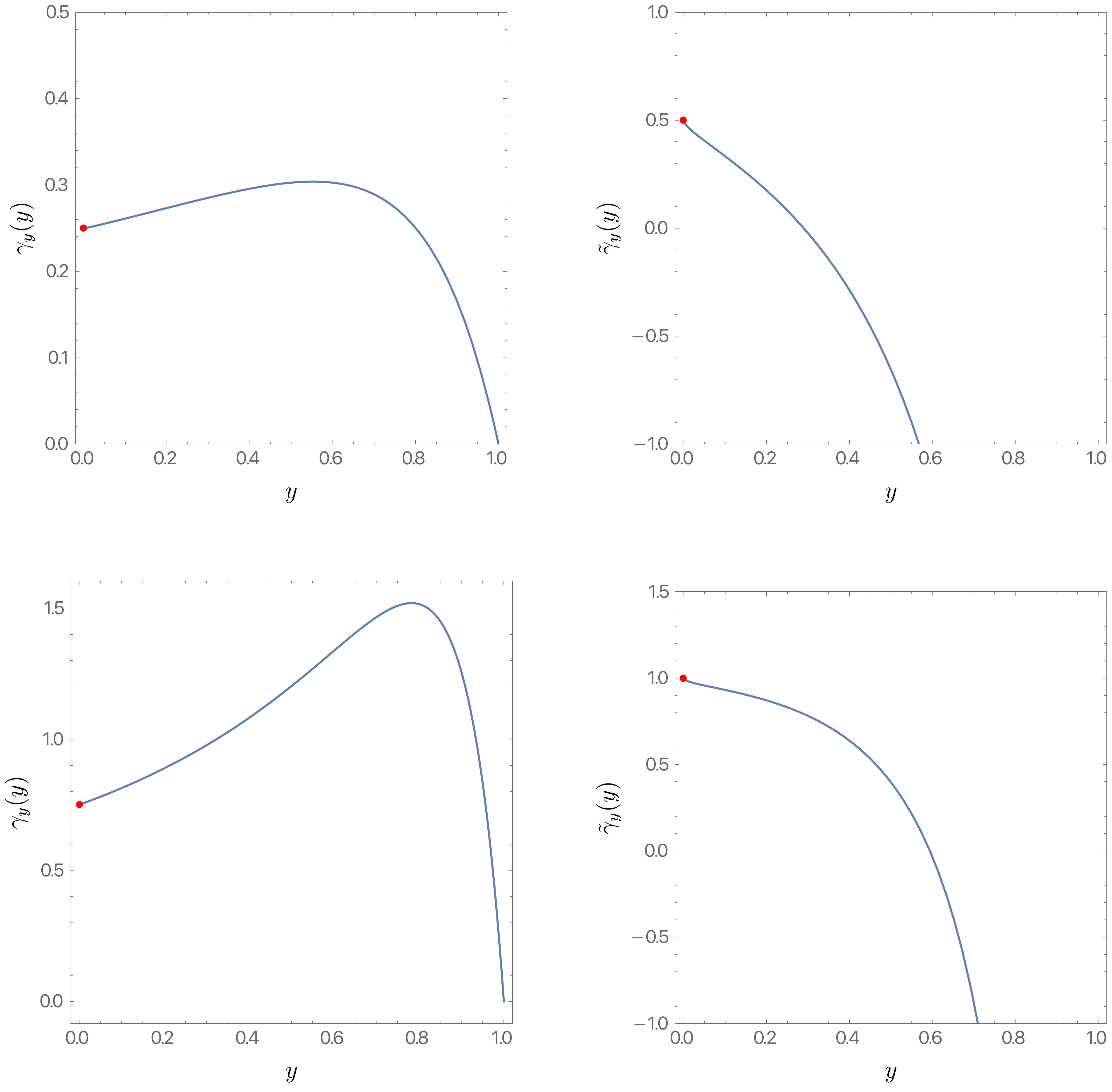}
\caption{\label{fig:simp}{\bf Left column}: Plots of $\gamma_y(y)$ as a function of $y$. {\bf Right column}: Plots of $\tilde{\gamma}_y(y)$ as a function of $y$. The top row has $\gamma=1/4$ and the bottom row has $\gamma=3/4$. In all cases, $\ell=1$ and $V=1$. The red dots denote the scaling exponents
derived from the linear analysis.}
\end{figure}

 After obtaining the exact solution, we extract the exponents by computing
 logarithmic derivatives  as a function of radius 
 \begin{equation}
 \gamma_y(y)\equiv \frac{y}{\phi}\frac{\mathrm{d}\phi}{\mathrm{d}y}\quad\text{and}\quad \tilde{\gamma}_y(y)\equiv \frac{y}{p-1}\frac{\mathrm{d}p}{\mathrm{d}y}
 \end{equation}
 If our linear analysis is accurate, we expect $\gamma_y(0)=\gamma$ and $\tilde{\gamma}_y(0)=2\gamma$ if $\gamma < 1/2$. If $\gamma > 1/2$, we expect $\tilde{\gamma}_y(0)=1$. In this case, even though the leading correction to the metric looks smooth, at the next order there are fractional powers which make the horizon singular.  The results are shown in Fig.~(\ref{fig:simp}) as a function of $y$ for $\ell=1$ and $V=1$. On the top row we have $\gamma=1/4$ ($y_+=3/\sqrt{10}$), whereas on the bottom row we have $\gamma=3/4$ ($y_+=\sqrt{11/126}$). On the left column we plot $\gamma_y(y)$, whereas on the right column we plot $\tilde{\gamma}_y(y)$. The numerical data is represented in blue, and the linear analytic prediction is given as a red disk at $y=0$.  The agreement between the numerical results and the analytic analysis is reassuring and shows that, despite starting the scalars with magnitude one asymptotically and having a  horizon singularity, the linear near horizon calculation is reproduced at the non-linear level. Note that our boundary conditions at no point assumed power law decay. Indeed, our boundary conditions assumed only that $\phi(r_+)=0$ and that $p(r_+)=q(r_+)=1$. In this sense, our nonlinear confirmation of the linear results is non-trivial.
 
 %%%%%%%%%%%%%%%%%%%%%%%%%%%%%%%%%%%
 \subsection{Non-zero temperature results\label{sec:finitescalars}}
 %%%%%%%%%%%%%%%%%%%%%%%%%%%%%%%%%%%
 Having discussed the zero temperature limit, we now turn our attention to the non-extremal case. In particular, in this section we aim to show that whenever the linear analysis predicts $\gamma<1/2$, the specific heat at constant charge $Q$, \emph{i.e.} $C_Q$, will show a leading \emph{anomalous scaling} in the small temperature expansion. In particular, according to the general arguments established in section \ref{sec:scaling} we expect the leading behaviour at small $T$ of the specific heat to be $C_Q\ \propto\ T^{2\gamma}$.
 
 We will again use the ingoing Eddington-Finkelstein coordinates of (\ref{eq:linesimp}). Since we want a non-degenerate horizon we demand that $f$ has a simple zero at $r=r_+$. Under this assumption, Eq.~(\ref{eq:fsimp}) and Eq.~(\ref{eq:phisimp}) develop a regular singular point at $r=r_+$. To implement the boundary conditions we define
 \begin{equation}
 f(r)=\left(1-\frac{r_+}{r}\right)\left(\frac{r^2}{L^2}+1+\frac{r_+ r}{L^2}+\frac{r_+^2}{L^2}-\frac{ Q^2}{r\,r_+}\right)q(r)\,.
 \end{equation}
Again, we introduce a compact coordinate as in Eq.~(\ref{eq:compact}), and regard $q$, $\phi$ and $p$ as functions of $y$, with $y=0$ being the location of the event horizon and $y=1$ the conformal boundary.

At the conformal boundary we demand $q(1)=p(1)=1$ and $\phi(1)=V$, just like we did for the extremal case. We are thus left with specifying the boundary conditions at the event horizon. Since $r=r_+$ is a regular singular point of Eq.~(\ref{eq:fsimp}) and Eq.~(\ref{eq:phisimp}), the boundary conditions for $q$ and $\phi$ follow from demanding regularity at $r=r_+$ and in particular yield
\begin{equation}
q(0) \left(\left.\frac{\mathrm{d}p}{\mathrm{d}y}\right|_{y=0}+2\right)=\left[2-\frac{y_+^2 \ell  (\ell +1) \phi (0)^2}{y_+^2+3 y_+^4-\tilde{Q}^2}\right]\quad\text{and}\quad \left.\frac{\mathrm{d}\phi}{\mathrm{d}y}\right|_{y=0}=\frac{y_+^2 \ell  (1+\ell ) \phi (0)}{\left(y_+^2+3 y_+^4-\tilde{Q}^2\right) q(0)}\,,
\end{equation}
where we defined $y_+\equiv r_+/L$ and $\tilde{Q}=Q/L$. For $p(y)$, we again demand $p(0)=1$.

Having determined the boundary conditions, we can now readily compute all quantities of interest. We are particularly interested in the behaviour of the specific heat at constant charge $C_Q$. This is determined via the usual thermodynamic relation
\begin{equation}
C_Q=T\left(\frac{\partial S}{\partial T}\right)_Q\,.
\end{equation}
In order to determine $C_Q$ as a function of $T$, we need to find the entropy $S$ and temperature $T$ of our novel black holes. These are given by
\begin{equation}
S = \frac{\pi\,y_+^2\,p(0)}{G}L^2\quad\text{and}\quad \tilde{T}\equiv L\,T = \frac{y_+^2(1+3 y_+^2)-\tilde{Q}^2}{4\pi  y_+^3} q(0)\,.
\end{equation}

The strategy is clear: we hold fixed a particular value of $\tilde{Q}$, and decrease $y_+$ thus decreasing the temperature.
The expected low temperature scaling depends on $\gamma$ which is uniquely determined by $\tilde Q$.  So we can predict $\gamma$ from the onset.
 It remains then to check whether $C_Q$ does exhibit the scaling predicted in section \ref{sec:scaling}. For a RN-AdS black hole, it is relatively easy to check that
\begin{equation}
\frac{G\,C_Q^{\rm RN}}{L^2}=\frac{2 \pi  y_+^2}{3 \tilde{Q}^2-y_+^2+3 y_+^4}\left[y_+^2\left(1+3 y_+^2\right)-\tilde{Q}^2\right]
\label{eq:speRNc}
\end{equation}
and thus near $\tilde{T}\approx 0$ we find
\begin{equation}
\frac{G\,C_Q^{\rm RN}}{L^2}\approx\frac{ \pi ^2}{3} \sqrt{\frac{2}{3}}(1+12 \tilde{Q}^2)^{1/4} \left(1-\frac{1}{\sqrt{1+12 \tilde{Q}^2}}\right)^{3/2}\;\tilde{T}+\mathcal{O}(\tilde{T}^2)
\label{eq:speRN}
\end{equation}

\begin{figure}[th]
\centering \includegraphics[width=0.9\textwidth]{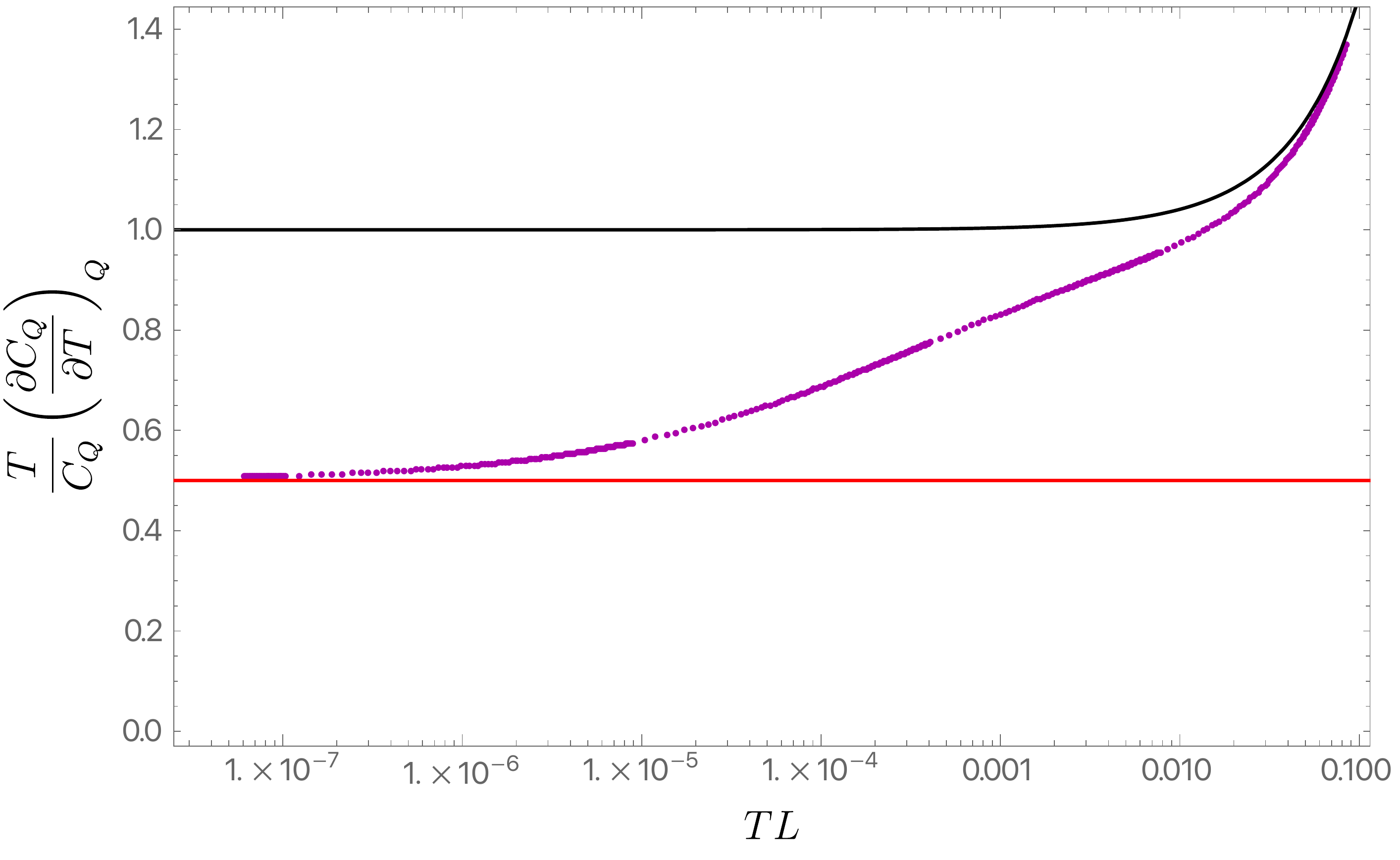}
\caption{\label{fig:specific_heat}Logarithmic derivative of $C_Q$, at constant charge $Q$, with respect to $T$. For the data shown we have used $\ell=1$ and ${Q/L}=3 \sqrt{37}/10$. The anomalous scaling predicted in section \ref{sec:scaling} is marked as a horizontal red line showing $2\gamma =1/2$, the purple disks are the exact numerical data collected at finite temperature and the solid black line shows the standard $AdS_2$ result of Eq.~(\ref{eq:speRNc}).}
\end{figure}

In Fig.~\ref{fig:specific_heat} we plot the logarithmic derivative of $C_Q$, at constant charge $Q$, with respect to $T$. This particular data was collected for $\ell=1$, $\gamma=1/4$, and thus $\tilde{Q}=3 \sqrt{37}/10$. Furthermore, we have chosen $V=1/2$. The purple disks are the numerical data, the solid black curve is given in Eq.~(\ref{eq:speRNc}) for the same charge $\tilde{Q}$ and the red horizontal line shows $1/2$ and is there to guide the eye. We can see that the logarithmic derivative approaches $2\gamma=1/2$ at low temperatures, as expected from our scaling in section \ref{sec:scaling}. For smaller values of $V$ we need to resort to (even) smaller temperatures to see the scaling emerging at small $T$. We have probed other values of $\tilde{Q}$ and find that whenever $\gamma<1/2$, we see an anomalous scaling. For $\gamma>1/2$, we return to the standard $AdS_2$ linear scaling given in (\ref{eq:speRN}). 
%%%%%%%%%%%%%%%%%%%%%%%%%%%%%%%%%%%%
\section{Discussion}
%%%%%%%%%%%%%%%%%%%%%%%%%%%%%%%%%%%%

%%%%%%%%%%%%%%%%%%%%%%%%%%%%%%%%%%%%
We have seen that as soon as one breaks the usual spatial symmetry, almost all stationary black holes in AdS develop curvature singularities on their horizon in the extremal limit. (The main exception is small toroidal black holes.) This singularity results in infinite tidal forces for infalling observers.
Contrary to the usual situation where the horizon curvature increases when the black hole becomes smaller, this singularity becomes stronger when the black hole becomes larger. Since the singularity arises in the limit of a family of smooth black holes, the solution is clearly physical \cite{Gubser:2000nd}, and not like the singularity of $M<0$ Schwarzschild. 

The diverging tidal forces close to the horizon should be proportional to the deviation from spherical symmetry at infinity.  Thus, one could be tempted to infer that (close to the spherical symmetry) the physical effects are negligibly small, despite the null singularity. However, to properly account for e.g. the {\it total} deformation an infalling observer experiences, one should look at the integrated tidal forces rather than their pointwise values. If the scaling dimension of the metric perturbation satisfies $\gamma > 1$, the forces are integrable, so the deformation would be finite and indeed small for the boundary conditions close to the spherical symmetry. However, for $\gamma <1$, the forces are not integrable. Thus, the total deformation remains infinite for arbitrary small deviations from the spherical symmetry. In particular, this is the case for RN AdS of any charge and for sufficiently large Kerr AdS black holes.

 Many supersymmetric solutions have been found in AdS with smooth horizons.\footnote{See \cite{Markeviciute:2018yal} for an example of a supersymmetric solution with a singular horizon.}  However, if one deforms the boundary conditions slightly, supersymmetry will be broken and we expect that the horizon will become singular. (We have checked this explicitly in one case.) So supersymmetric solutions are very special, like Reissner-Nordstr\"om AdS, and do not see these singularities. 
 
 The Einstein-Maxwell theory we have studied can be embedded in supergravity in different ways. In some embeddings, there are charged scalars with mass $m$ and charge $q$. Depending on ($m,q$), there may be a range of $r_+$ in which the extremal black hole is unstable to turning on the scalar field \cite{Dias:2016pma}. In this case, the singularity is likely to become much worse \cite{Horowitz:2009ij}. However, even with charged scalars, there is usually a range of $r_+$ for which there is no instability to turning on the scalar field, and in those cases, our tidal force singularity will remain. (An extensive discussion of the near horizon scaling dimensions of supergravity fields is given in \cite{Castro:2021wzn}.)

It is natural to ask how stringy or quantum effects will modify this singularity.  Although we cannot answer this, we make the following comments. Infalling strings will certainly become excited by the large tidal forces and should backreact on the geometry. Quantum effects are often discussed in the context of the Euclidean theory. All of the examples we have considered, except for the rotating black holes in Sec. 3.3, are static and have a real Euclidean analytic continuation.\footnote{The Maxwell field will also remain real if we consider the magnetic rather than electrically charged solutions. Since the four dimensional Maxwell stress tensor is symmetric with respect to exchanging magnetic and electric field, the metric remains the same.} In the extremal limit, the horizon moves off to infinity and becomes another asymptotic boundary to the Euclidean solution. All curvature invariants of the Euclidean solution are the same as the Lorentzian solution. Since the latter remain finite, and the Euclidean curvature is completely determined by its scalar invariants,  the Euclidean solution is completely nonsingular! The only remnant of the tidal force singularity seen in the Lorentzian solution is that the solution decays with noninteger powers of the radius in the asymptotic region associated with the horizon. But that also occurs near the boundary at infinity for most massive fields in AdS, and does not cause any problems. Thus there is no reason to discard these solutions.
In particular, they are proper saddles for the Euclidean path integral with appropriate boundary conditions. It follows, that such black holes may be also prepared (as a quantum state) using the Euclidean path integral.
 
 As we have discussed, when these tidal force singularities are strong enough, there is a clear signal of them in a holographic dual theory. Various quantities such as the specific heat will exhibit anomalous scaling with temperature as $T\to 0$. For topologically spherical, charged black holes, this applies whenever the black hole is larger than roughly the AdS radius. We have also considered two examples of extreme black holes in pure gravity: Kerr AdS and hyperbolic black holes. These black holes exhibit more mild singularities which holographically appear in anomalous higher order corrections to the leading (linear in $T$) behavior of the specific heat.

Finally, we mention a holographic argument that a singularity should perhaps be expected on the horizon of an extremal black hole and not be resolved by quantum effects.\footnote{We thank Leonel Queimada for suggesting this.}  The entanglement wedge of the entire boundary of an extremal black hole only covers the region of spacetime outside the horizon. Since there is no other boundary, it appears that the region inside the horizon cannot be described in terms of the dual theory. If the theories are really equivalent, then perhaps spacetime should end at a singular horizon. Note that this argument does not depend on having a timelike singularity inside the black hole. The event horizon is a Cauchy horizon for any complete spacelike surface outside the black hole.
%%%%%%%%%%%%%%%%%%%%%%%%%%%%%%%%%%%%

%%%%%%%%%%%%%%%%%%%%%%%%
\subsection*{Acknowledgments}

It is a pleasure to thank Jan Boruch, Sean Hartnoll, Jerzy Lewandowski, and Don Marolf for discussions.
The work of G.~H. was supported in part by NSF Grant PHY-2107939. J.~E.~S. has been partially supported by STFC consolidated grant ST/T000694/1. M.~K. thanks University of California in Santa Barbara for the hospitality, he was partially supported by The Polish-U.S. Fulbright Commission and from Polish budgetary funds for science in 2018-2022 as a research project under the program "Diamentowy Grant".

\appendix
\section{Anomalous scaling of the specific heat}
%%%%%%%%%%%%%%%%%%%%%%%%%%%%%%%%%%%%

In this Appendix, we provide some details of the calculation of the anomalous scaling of the specific heat with temperature. The final answer agrees with  the scaling argument \eqref{anomcq} given in Sec. \ref{sec:scaling}.
We will  use second order perturbation theory about the near horizon geometry of an extreme RN-AdS black hole.

Since our argument relies on second order perturbation theory, our calculations will become less explicit and, for the sake of presentation, less general. We will focus on deformations of the near horizon geometry that break $SO(3)$ but preserve a $U(1)_{\phi}$ symmetry. Additionally we will impose the discrete symmetry $\phi\to-\phi$. We then write an Ans\"atze for the deformed metric and gauge field in Bondi-Sachs coordinates adapted to the extremal horizon:
\begin{subequations}
\begin{multline}
\mathrm{d}s^2=L^2\Bigg\{-A(\rho,\theta)\,\rho^2\,\frac{\mathrm{d}v^2}{L^2}+\frac{2\,\mathrm{d}v\,\mathrm{d}\rho}{L}
\\
+y_+^2\,H_L(\rho,\theta)\left[H_T(\rho,\theta)\left(\mathrm{d}\theta-\frac{U_\theta(\rho,\theta)\,\rho\,\mathrm{d}v}{L}\right)^2+\frac{\sin^2\theta}{H_T(\rho,\theta)}\mathrm{d}\phi^2\right]\Bigg\}\,,
\end{multline}
and
\begin{equation}
A=\rho\,A_v(\rho,\theta)\,\mathrm{d}v+L\,A_\theta(\rho,\theta)\,\mathrm{d}\theta\,,
\end{equation}
\end{subequations}
with the extremal horizon being located at $\rho=0$, as usual.

We now expand all metric and gauge field functions in terms of harmonics on the round $S^2$, by identifying how each transforms under diffeomorphism on the two-sphere. This will make  contact with section \ref{sec:per}. Essentially, $A$, $H_L$ and $A_v$ can be written as an infinite sum of scalar spherical harmonics and $U_\theta$ and $A_\theta$ as an infinite sum of gradients of spherical harmonics. $H_T$ is slightly more subtle, but its transformations properties can be read from (\ref{eq:ttpiece}). Note that no spherical vector harmonics appear in our expansion, since these would necessarily break the discrete symmetry $\phi\to-\phi$ that we want to preserve. To sum up, we have
\begin{subequations}
\begin{equation}
A(\rho,\theta)=\sum_{\ell=0}^{+\infty} \mathbb{S}_{\ell}(\theta)a^{\ell}(\rho)\,,
\end{equation}

\begin{equation}
H_L(\rho,\theta)=\sum_{\ell=0}^{+\infty} \mathbb{S}_{\ell}(\theta)h_L^{\ell}(\rho)\,,
\end{equation}
\begin{equation}
A_v(\rho,\theta)=\sum_{\ell=0}^{+\infty} \mathbb{S}_{\ell}(\theta)a_v^{\ell}(\rho)\,,
\end{equation}
\begin{equation}
U_\theta(\rho,\theta)=\sum_{\ell=1}^{+\infty} \partial_{\theta}\mathbb{S}_{\ell}(\theta)u_\theta^{\ell}(\rho)\,,
\end{equation}
\begin{equation}
A_{\theta}(\rho,\theta)=\sum_{\ell=1}^{+\infty} \partial_{\theta}\mathbb{S}_{\ell}(\theta)a_{\theta}^{\ell}(\rho)\,,
\end{equation}
and
\begin{equation}
H_T(\rho,\theta)=\sum_{\ell=2}^{+\infty} \left[\ell(\ell+1)\mathbb{S}_{\ell}+2 \cot \theta\, \partial_{\theta}\mathbb{S}_{\ell}(\theta)\right]h_T^{\ell}(\rho)\,.
\end{equation}
\end{subequations}%%%
So far we have not made any approximation. Indeed, we could have used this expansion to perform the full nonlinear numerical analysis of section (\ref{sec:num}), precisely in the spirit of Galerkin spectral methods.

We now introduce our approximation scheme. We expand each of the functions $\{a^\ell,h_L^\ell,a_v^{\ell},u_\theta^{\ell},a_{\theta}^{\ell},h_T^\ell\}$ as a power series in a book keeping parameter $\varepsilon$ which we take to be small
\begin{subequations}
\begin{equation}
a^\ell(\rho)=\sum_{j=0}^{+\infty}\varepsilon^j a^{\ell}_{(j)}(\rho)\,,
\end{equation}
\begin{equation}
h^\ell_L(\rho)=\sum_{j=0}^{+\infty}\varepsilon^j h_{L\,(j)}^{\ell}(\rho)\,,
\end{equation}
\begin{equation}
a^\ell_v(\rho)=\sum_{j=0}^{+\infty}\varepsilon^j a_{v\,(j)}^{\ell}(\rho)\,,
\end{equation}
\begin{equation}
u^\ell_\theta(\rho)=\sum_{j=0}^{+\infty}\varepsilon^j u_{\theta\,(j)}^{\ell}(\rho)\,,
\end{equation}
\begin{equation}
a^\ell_\theta(\rho)=\sum_{j=0}^{+\infty}\varepsilon^j a_{\theta\,(j)}^{\ell}(\rho)\,,
\end{equation}
and
\begin{equation}
h^\ell_T(\rho)=\sum_{j=0}^{+\infty}\varepsilon^j h_{T\,(j)}^{\ell}(\rho)\,.
\end{equation}
\label{eqs:lots}
\end{subequations}

We take the background, \emph{i.e.} the order $\varepsilon^0$, to be given by an extreme RN-AdS black hole, which amounts to taking
\begin{equation}
a^0_{(0)}(\rho)=6+\frac{1}{y_+^2}\,,\quad h_{L\,(0)}^0=1\quad\text{and}\quad a_{v\,(0)}^0=\frac{\sqrt{1+3 y_+^2}}{y_+}\,.
\end{equation}
with all the remaining coefficients in (\ref{eqs:lots}) set to zero.

At linear order, \emph{i.e.} $\varepsilon^1$, we impose an $\ell=2$ deformation. If we imagine that these near horizon deformations arise from a boundary deformation of a generic profile, the $\ell=2$ perturbation is the one that decays the slowest as we approach the horizon, and as such provides the leading effect we want to study. In order to only keep a $\ell=2$ deformation, we take $a_{(1)}^2(\rho)$, $u_{\theta\,(1)}^2(\rho)$, $h^2_{L\,(1)}(\rho)$, $h^2_{T\,(1)}(\rho)$, $a^2_{v\,(1)}(\rho)$ and $a^2_{\theta\,(1)}(\rho)$ to be non-vanishing, but keep all the remaining coefficients with $\ell>2$ zero. Solving the linear Einstein-Maxwell equations yields
\begin{subequations}
\begin{align}
a_{(1)}^2(\rho)&=\rho^\gamma\,,
\\
u^2_{\theta\,(1)}(\rho)&=\frac{1}{\left(1+6 y_+^2\right) (\gamma -1)}\,\rho^{\gamma}\,,
\\
h^2_{L\,(1)}(\rho)&=0\,,
\\
h_{T\,(1)}^2(\rho)&=\frac{y_+^2 (\gamma +1)}{4 \left(1+6 y_+^2\right)}\frac{\gamma ^2+\gamma -10+6 y_+^2 \left(\gamma ^2+\gamma -2\right)}{6+\gamma  \left[\gamma ^2-7+6 y_+^2 \left(\gamma ^2-1\right)\right]}\,\rho^{\gamma}\,,
\\
a_{v\,(1)}^2(\rho)&=\frac{6 y_+}{1+6 y_+^2}\frac{\sqrt{1+3y_+^2}}{6+\gamma  \left[\gamma ^2-7+6 y_+^2 \left(\gamma ^2-1\right)\right]}\,\rho^{\gamma}\,,
\\
a_{\theta\,(1)}^2(\rho)&=\frac{y_+^3}{1+6 y_+^2}\frac{\sqrt{1+3y_+^2}}{6+\gamma  \left[\gamma ^2-7+6 y_+^2 \left(\gamma ^2-1\right)\right]}\,\rho^{\gamma}\,,
\end{align}
where the first equation defines what we mean by $\varepsilon$. The exponent $\gamma$ can again take four distinct values. Two are negative, and we ignore those via boundary conditions, and we are interested in the smallest of the two that are positive. This yields
\begin{equation}
\gamma = \frac{1}{2} \left[\sqrt{5+\frac{24}{1+6 y_+^2}-4 \sqrt{1+\frac{24 \left(1+3 y_+^2\right)}{\left(1+6 y_+^2\right){}^2}}}-1\right]\,,
\label{eq:direct}
\end{equation}
which matches $\gamma_{+-}$ in Eq.~(\ref{n_spherical}) with $\ell=2$, as it should. Note that $h^2_{L\,(1)}$ vanishes via the equations of motion.
\end{subequations}

One can now proceed to second order in $\varepsilon$. At quadratic order, modes with $\ell=0,2,4$ are generated in the expansion. The final expression for each of the coefficients is rather complicated and not very illuminating. However, in order to make the argument we want, we only need to focus on $H_L$. The reason for this is that the area of a surface of constant $v$ and $\rho=\rho_0$ is simply given by
\begin{equation}
A(\rho_0)=2\pi L^2\,y_+^2\int_0^{\pi}H_L(\rho_0,\theta)\,\sin\theta\,\mathrm{d}\theta=4\pi L^2\,y_+^2 h^0_L(\rho_0)\,,
\label{eq:generalentropy}
\end{equation}
where we used that spherical harmonics have vanishing integral over the sphere. Note that the above expression is exact. From this expression, it is clear that we can only get contributions to the area coming from the $\ell=0$ harmonic. This justifies why we need to go to second order in $\varepsilon$ to see the effect we want.

After some algebra, one finds
\begin{subequations}
\begin{equation}
h^0_{L\,(2)}(\rho)=C_2(\gamma)\,\rho^{2\gamma}+C_0+C_1\,\rho\,,
\label{eq:simpleh0}
\end{equation}
with
\begin{multline}
C_2(\gamma)=\frac{(\gamma +1) (\gamma +2)}{5760 (1-2 \gamma ) (1-\gamma ) \left(\gamma ^2+\gamma +6\right)}\Bigg[48+12 \gamma -4 \gamma ^2-35 \gamma ^3-25 \gamma ^4-9 \gamma ^5
\\
-3 \gamma ^6+\frac{72-22 \gamma -27 \gamma ^2-10 \gamma ^3-5 \gamma ^4}{\sqrt{3}}\sqrt{\gamma  (\gamma +1) \left(\gamma ^2+\gamma +4\right)}\Bigg]
\end{multline}
\label{eq:beast}
\end{subequations}
where we regard $y_+$ as a function of $\gamma$ by inverting (\ref{eq:direct}), \emph{i.e.}
\begin{equation}
y_+=\frac{1}{\sqrt{6}}\frac{1}{ \sqrt{(1-\gamma ) (\gamma +2)}}\sqrt{\gamma  (\gamma +1)-8+2 \sqrt{3+\frac{12}{\gamma  (\gamma +1)}}}\,.
\end{equation}
In the above, $C_0$ and $C_1$ are integration constants. We set $C_0=0$, which essentially defines $y_+$. $C_1$, on the other hand, \emph{cannot} be set to zero.

The exact form of Eq.~(\ref{eq:beast}) is largely unimportant, except for a few of points. The most important point is that it turns out to be non-zero, unlike at first order in $\varepsilon$. The second important point is that in the range $0<\gamma<1/2$, the leading contribution to $h^0_{L\,(2)}(\rho)$ is proportional to $\rho^{2\gamma}$.  Furthermore, its coefficient, $C_2(\gamma)$, turns out to be positive definite (see Fig.~\ref{fig:coef} below).
\begin{figure}[th]
\centering \includegraphics[width=0.6\textwidth]{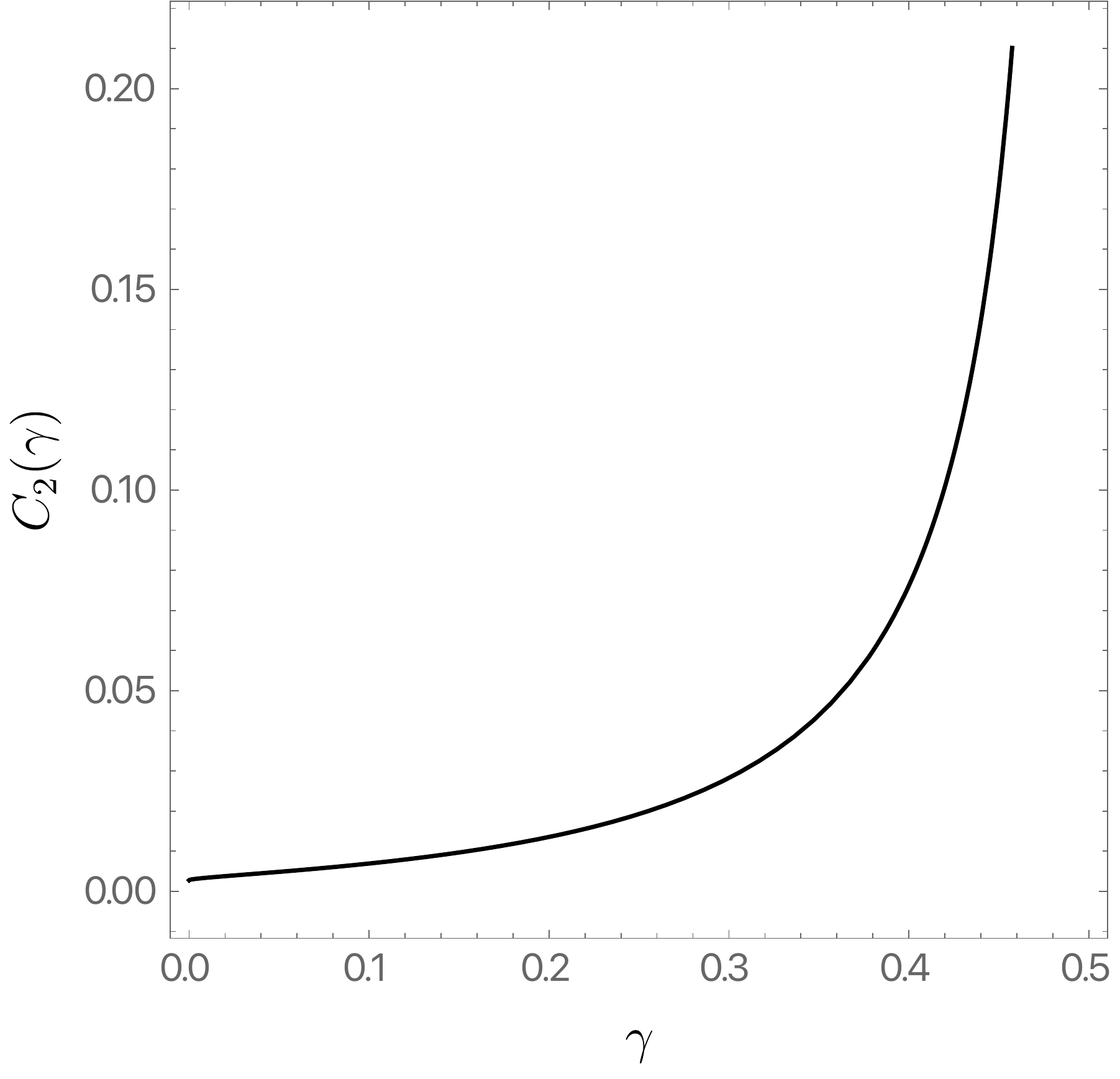}
\caption{\label{fig:coef} $C_2(\gamma)$ as a function of $\gamma$ in the range $0<\gamma<1/2$.}
\end{figure}

The apparent singularity at $\gamma=1/2$ reflects the fact that, for that value of $\gamma$, $h^0_{L\,(2)}(\rho)$ is no longer given as in (\ref{eq:simpleh0}), but instead
\begin{equation}
h^0_{L\,(2)}=C_0+C_1 \rho +\frac{997+281 \sqrt{19}}{110592} \rho\,\log \left(\frac{1}{\rho}\right)\,,
\end{equation}
where again we can set $C_0=0$, but not $C_1$. The leading behaviour near $\rho=0$ is then given by $\rho \log (1/\rho)$, again with a positive coefficient.

We can now apply the same scaling argument we used in section \ref{sec:scaling}  to the black hole entropy. Indeed, if we use (\ref{eq:generalentropy}), we have just shown that a surface of constant $v$ and $\rho=\rho_0$, to leading order in $\rho_0\ll1$, has an area given by
\begin{equation}
A(\rho_0) = 4\pi L^2 y_+^2\left[1+C_2(\gamma)\;\rho_0^{2\gamma}\right]\,,
\end{equation}
in the range $0<\gamma<1/2$, which yields an entropy of the black hole horizon scaling as
\begin{equation}
S \approx S_0+S_2\,T^{2\gamma}
\end{equation}
where $S_0$ and $S_2$ are suitable constants. Using similar arguments, applied for $\gamma=1/2$, we find instead
\begin{equation}
S \approx \tilde{S}_0+\tilde{S}_2\,T\,\log T+\tilde{S}_3 T\,,
\end{equation}
for suitable constants $\tilde{S}_0$, $\tilde{S}_2$ and $\tilde{S}_3$. Note that $\gamma$ is ultimately fixed by the total charge $Q$ (or alternatively $y_+$). Using standard thermodynamic relations, we predict the scaling of the specific heat at constant charge $Q$, at sufficiently small temperatures, to be given by
\begin{equation}
C_Q \approx 2\gamma\tilde{S}_2 T^{2\gamma}
\label{eq:scaling}
\end{equation}
in the range $0<\gamma<1/2$ and 
\begin{equation}
C_Q \approx T \left(\tilde{S}_2+\tilde{S}_3\right)+\tilde{S}_2 T \log T\,,
\end{equation}
for $\gamma=1/2$. Note that the fact that $C_2(\gamma)$ is positive is paramount to argue that the near horizon geometry we found is thermodynamically locally stable, since $\tilde{S}_2$ turns out to be proportional to $C_2(\gamma)$, which makes $C_Q$ in (\ref{eq:scaling}) also positive.

Though we have deduced this anomalous scaling for Einstein-Mawell, we predict this will be true for a variety of systems that suffer from tidal force-type singularities. For instance, in section \ref{sec:finitescalars} we  show that our finite temperature scalar model exhibits a scaling of the form (\ref{eq:scaling}).

\bibliographystyle{JHEP}
\bibliography{bibl}

\providecommand{\href}[2]{#2}\begingroup\raggedright\begin{thebibliography}{10}

\bibitem{Welch:1995dh}
D.L.~Welch, \emph{{On the smoothness of the horizons of multi - black hole
  solutions}}, \href{https://doi.org/10.1103/PhysRevD.52.985}{\emph{Phys. Rev.
  D} {\bfseries 52} (1995) 985}
  [\href{https://arxiv.org/abs/hep-th/9502146}{{\ttfamily hep-th/9502146}}].

\bibitem{Candlish:2007fh}
G.N.~Candlish and H.S.~Reall, \emph{{On the smoothness of static multi-black
  hole solutions of higher-dimensional Einstein-Maxwell theory}},
  \href{https://doi.org/10.1088/0264-9381/24/23/022}{\emph{Class. Quant. Grav.}
  {\bfseries 24} (2007) 6025}
  [\href{https://arxiv.org/abs/0707.4420}{{\ttfamily 0707.4420}}].

\bibitem{Dias:2011at}
O.J.C.~Dias, G.T.~Horowitz and J.E.~Santos, \emph{{Black holes with only one
  Killing field}}, \href{https://doi.org/10.1007/JHEP07(2011)115}{\emph{JHEP}
  {\bfseries 07} (2011) 115} [\href{https://arxiv.org/abs/1105.4167}{{\ttfamily
  1105.4167}}].

\bibitem{Maeda:2011pk}
K.~Maeda, T.~Okamura and J.-i.~Koga, \emph{{Inhomogeneous charged black hole
  solutions in asymptotically anti-de Sitter spacetime}},
  \href{https://doi.org/10.1103/PhysRevD.85.066003}{\emph{Phys. Rev. D}
  {\bfseries 85} (2012) 066003}
  [\href{https://arxiv.org/abs/1107.3677}{{\ttfamily 1107.3677}}].

\bibitem{Hickling:2015ooa}
A.~Hickling, \emph{{Bulk Duals for Generic Static, Scale-Invariant Holographic
  CFT States}},
  \href{https://doi.org/10.1088/0264-9381/32/17/175011}{\emph{Class. Quant.
  Grav.} {\bfseries 32} (2015) 175011}
  [\href{https://arxiv.org/abs/1504.03723}{{\ttfamily 1504.03723}}].

\bibitem{Iizuka:2022igv}
N.~Iizuka, A.~Ishibashi and K.~Maeda, \emph{{Flows of extremal attractor black
  holes}}, \href{https://doi.org/10.1007/JHEP09(2022)093}{\emph{JHEP}
  {\bfseries 09} (2022) 093}
  [\href{https://arxiv.org/abs/2206.04845}{{\ttfamily 2206.04845}}].

\bibitem{Markeviciute:2018yal}
J.~Markeviciute and J.E.~Santos, \emph{{Evidence for the existence of a novel
  class of supersymmetric black holes with AdS$_5\times$S$^5$ asymptotics}},
  \href{https://doi.org/10.1088/1361-6382/aaf680}{\emph{Class. Quant. Grav.}
  {\bfseries 36} (2019) 02LT01}
  [\href{https://arxiv.org/abs/1806.01849}{{\ttfamily 1806.01849}}].

\bibitem{Horowitz:2022leb}
G.T.~Horowitz, M.~Kolanowski and J.E.~Santos, \emph{{A deformed IR: a new IR
  fixed point for four-dimensional holographic theories}},
  \href{https://arxiv.org/abs/2211.01385}{{\ttfamily 2211.01385}}.

\bibitem{Kunduri:2008tk}
H.K.~Kunduri and J.~Lucietti, \emph{{Uniqueness of near-horizon geometries of
  rotating extremal AdS(4) black holes}},
  \href{https://doi.org/10.1088/0264-9381/26/5/055019}{\emph{Class. Quant.
  Grav.} {\bfseries 26} (2009) 055019}
  [\href{https://arxiv.org/abs/0812.1576}{{\ttfamily 0812.1576}}].

\bibitem{Aretakis:2011ha}
S.~Aretakis, \emph{{Stability and Instability of Extreme Reissner-Nordstr\"om
  Black Hole Spacetimes for Linear Scalar Perturbations I}},
  \href{https://doi.org/10.1007/s00220-011-1254-5}{\emph{Commun. Math. Phys.}
  {\bfseries 307} (2011) 17} [\href{https://arxiv.org/abs/1110.2007}{{\ttfamily
  1110.2007}}].

\bibitem{Horowitz:2014gva}
G.T.~Horowitz, N.~Iqbal, J.E.~Santos and B.~Way, \emph{{Hovering Black Holes
  from Charged Defects}},
  \href{https://doi.org/10.1088/0264-9381/32/10/105001}{\emph{Class. Quant.
  Grav.} {\bfseries 32} (2015) 105001}
  [\href{https://arxiv.org/abs/1412.1830}{{\ttfamily 1412.1830}}].

\bibitem{Li:2015wsa}
C.~Li and J.~Lucietti, \emph{{Transverse deformations of extreme horizons}},
  \href{https://doi.org/10.1088/0264-9381/33/7/075015}{\emph{Class. Quant.
  Grav.} {\bfseries 33} (2016) 075015}
  [\href{https://arxiv.org/abs/1509.03469}{{\ttfamily 1509.03469}}].

\bibitem{Fontanella:2016lzo}
A.~Fontanella and J.B.~Gutowski, \emph{{Moduli Spaces of Transverse
  Deformations of Near-Horizon Geometries}},
  \href{https://doi.org/10.1088/1751-8121/aa6cbf}{\emph{J. Phys. A} {\bfseries
  50} (2017) 215202} [\href{https://arxiv.org/abs/1610.09949}{{\ttfamily
  1610.09949}}].

\bibitem{Li:2018knr}
C.~Li and J.~Lucietti, \emph{{Electrovacuum spacetime near an extreme
  horizon}}, \href{https://doi.org/10.4310/ATMP.2019.v23.n7.a5}{\emph{Adv.
  Theor. Math. Phys.} {\bfseries 23} (2019) 1903}
  [\href{https://arxiv.org/abs/1809.08164}{{\ttfamily 1809.08164}}].

\bibitem{Kolanowski:2019wua}
M.~Kolanowski, J.~Lewandowski and A.~Szereszewski, \emph{{Extremal horizons
  stationary to the second order: new constraints}},
  \href{https://doi.org/10.1103/PhysRevD.100.104057}{\emph{Phys. Rev. D}
  {\bfseries 100} (2019) 104057}
  [\href{https://arxiv.org/abs/1907.00955}{{\ttfamily 1907.00955}}].

\bibitem{Kolanowski:2021tje}
M.~Kolanowski, \emph{{Towards the black hole uniqueness: transverse
  deformations of the extremal Reissner-Nordstr\"om-(A)dS horizon}},
  \href{https://doi.org/10.1007/JHEP01(2022)042}{\emph{JHEP} {\bfseries 01}
  (2022) 042} [\href{https://arxiv.org/abs/2111.00806}{{\ttfamily
  2111.00806}}].

\bibitem{yang1980eigenvalues}
P.C.~Yang and S.-T.~Yau, \emph{Eigenvalues of the laplacian of compact riemann
  surfaces and minimal submanifolds}, {\emph{Annali della Scuola Normale
  Superiore di Pisa-Classe di Scienze} {\bfseries 7} (1980) 55}.

\bibitem{el1983volume}
A.~El~Soufi and S.~Ilias, \emph{Le volume conforme et ses applications
  d'apr{\`e}s li et yau}, {\emph{S{\'e}minaire de th{\'e}orie spectrale et
  g{\'e}om{\'e}trie} {\bfseries 2} (1983) 1}.

\bibitem{Kravchuk:2021akc}
P.~Kravchuk, D.~Mazac and S.~Pal, \emph{{Automorphic Spectra and the Conformal
  Bootstrap}},  \href{https://arxiv.org/abs/2111.12716}{{\ttfamily
  2111.12716}}.

\bibitem{Bonifacio:2021aqf}
J.~Bonifacio, \emph{{Bootstrapping closed hyperbolic surfaces}},
  \href{https://doi.org/10.1007/JHEP03(2022)093}{\emph{JHEP} {\bfseries 03}
  (2022) 093} [\href{https://arxiv.org/abs/2111.13215}{{\ttfamily
  2111.13215}}].

\bibitem{Dias:2012pp}
O.J.C.~Dias, J.E.~Santos and M.~Stein, \emph{{Kerr-AdS and its Near-horizon
  Geometry: Perturbations and the Kerr/CFT Correspondence}},
  \href{https://doi.org/10.1007/JHEP10(2012)182}{\emph{JHEP} {\bfseries 10}
  (2012) 182} [\href{https://arxiv.org/abs/1208.3322}{{\ttfamily 1208.3322}}].

\bibitem{Dias:2015rxy}
O.J.C.~Dias, J.E.~Santos and B.~Way, \emph{{Black holes with a single Killing
  vector field: black resonators}},
  \href{https://doi.org/10.1007/JHEP12(2015)171}{\emph{JHEP} {\bfseries 12}
  (2015) 171} [\href{https://arxiv.org/abs/1505.04793}{{\ttfamily
  1505.04793}}].

\bibitem{Chesler:2013qla}
P.~Chesler, A.~Lucas and S.~Sachdev, \emph{{Conformal field theories in a
  periodic potential: results from holography and field theory}},
  \href{https://doi.org/10.1103/PhysRevD.89.026005}{\emph{Phys. Rev. D}
  {\bfseries 89} (2014) 026005}
  [\href{https://arxiv.org/abs/1308.0329}{{\ttfamily 1308.0329}}].

\bibitem{Biggs:2022lvi}
W.D.~Biggs and J.E.~Santos, \emph{{Black Tunnels and Hammocks}},
  \href{https://arxiv.org/abs/2207.14306}{{\ttfamily 2207.14306}}.

\bibitem{Dias:2015nua}
O.J.C.~Dias, J.E.~Santos and B.~Way, \emph{{Numerical Methods for Finding
  Stationary Gravitational Solutions}},
  \href{https://doi.org/10.1088/0264-9381/33/13/133001}{\emph{Class. Quant.
  Grav.} {\bfseries 33} (2016) 133001}
  [\href{https://arxiv.org/abs/1510.02804}{{\ttfamily 1510.02804}}].

\bibitem{Gubser:2000nd}
S.S.~Gubser, \emph{{Curvature singularities: The Good, the bad, and the
  naked}}, \href{https://doi.org/10.4310/ATMP.2000.v4.n3.a6}{\emph{Adv. Theor.
  Math. Phys.} {\bfseries 4} (2000) 679}
  [\href{https://arxiv.org/abs/hep-th/0002160}{{\ttfamily hep-th/0002160}}].

\bibitem{Dias:2016pma}
O.J.C.~Dias and R.~Masachs, \emph{{Hairy black holes and the endpoint of
  AdS$_4$ charged superradiance}},
  \href{https://doi.org/10.1007/JHEP02(2017)128}{\emph{JHEP} {\bfseries 02}
  (2017) 128} [\href{https://arxiv.org/abs/1610.03496}{{\ttfamily
  1610.03496}}].

\bibitem{Horowitz:2009ij}
G.T.~Horowitz and M.M.~Roberts, \emph{{Zero Temperature Limit of Holographic
  Superconductors}},
  \href{https://doi.org/10.1088/1126-6708/2009/11/015}{\emph{JHEP} {\bfseries
  11} (2009) 015} [\href{https://arxiv.org/abs/0908.3677}{{\ttfamily
  0908.3677}}].

\bibitem{Castro:2021wzn}
A.~Castro and E.~Verheijden, \emph{{Near-AdS2 Spectroscopy: Classifying the
  Spectrum of Operators and Interactions in N=2 4D Supergravity}},
  \href{https://doi.org/10.3390/universe7120475}{\emph{Universe} {\bfseries 7}
  (2021) 475} [\href{https://arxiv.org/abs/2110.04208}{{\ttfamily
  2110.04208}}].

\end{thebibliography}\endgroup



\providecommand{\href}[2]{#2}\begingroup\raggedright\endgroup
\end{document}